\documentclass[twocolumn,aps,prd,amsmath,amssymb,floatfix,amsmath,superscriptaddress]{revtex4-2}

\usepackage{commath}
\usepackage{units}
\usepackage{graphicx}
\usepackage{dcolumn}
\usepackage{bm}
\usepackage{braket}
\usepackage{color,epsfig}
\usepackage{slashed}
\usepackage{hyperref}
\usepackage{subfigure}
\usepackage{feynmf}
\usepackage{calrsfs}

\usepackage{booktabs}

\newcommand{\bea}{\begin{eqnarray}}
\newcommand{\eea}{\end{eqnarray}}
\newcommand{\be}{\begin{equation}}
\newcommand{\ee}{\end{equation}}

\renewcommand\vec{\bm}

\DeclareMathOperator\arctanh{arctanh}

\DeclareFontFamily{U}{calligra}{}
\DeclareFontShape{U}{calligra}{m}{n}{<->callig15}{}

\newcommand{\nn}{\nonumber}

\DeclareMathAlphabet{\mathcal}{OMS}{cmsy}{m}{n}

\begin{document}


\title{Very Special Relativity in Accelerated Frames:\\
Non-relativistic Effects in Gravitational Spectroscopy of Ultracold Neutrons}


\author{Alessandro Santoni}
\email{asantoni@uc.cl}
\affiliation{Institut f\"ur Theoretische Physik and Atominstitut,
 Technische Universit\"at Wien,
 Wiedner Hauptstrasse 8--10,
 A-1040 Vienna, Austria}
 \affiliation{Facultad de F\'isica, Pontificia Universidad Cat\'olica de Chile, Vicu\~{n}a Mackenna 4860, Santiago, Chile}
 
\author{Enrique Mu\~{n}oz}
\email{ejmunozt@uc.cl}
\affiliation{Facultad de F\'isica, Pontificia Universidad Cat\'olica de Chile, Vicu\~{n}a Mackenna 4860, Santiago, Chile}

\author{Hartmut Abele}
 \email{hartmut.abele@tuwien.ac.at}
\affiliation{Atominstitut TU Wien, Stadionallee 2, 1020 Vienna, Austria}

\author{Benjamin Koch}
 \email{benjamin.koch@tuwien.ac.at}
\affiliation{Institut f\"ur Theoretische Physik and Atominstitut,
 Technische Universit\"at Wien,
 Wiedner Hauptstrasse 8--10,
 A-1040 Vienna, Austria}
\affiliation{Facultad de F\'isica, Pontificia Universidad Cat\'olica de Chile, Vicu\~{n}a Mackenna 4860, Santiago, Chile}


\date{\today}

\begin{abstract}

 In this paper, we investigate the phenomenology of fermionic systems in uniform gravitational fields within the framework of very special relativity (VSR). We focus on the case of gravitational spectroscopy with ultracold neutrons, explored in experiments like \emph{q}\textsc{Bounce}. Calculating the leading ($c^0$) and next-to-leading ($c^{-1}$) order corrections to the non-relativistic Hamiltonian in an accelerated frame, we obtain the perturbed fermionic energy spectrum. At leading order, we do not find any modifications except for a trivial mass shift, thus preserving the equivalence between inertial and gravitational mass and particle-antiparticle sectors. The next-to-leading order corrections, instead, introduce time-dependent anisotropic contributions depending on the preferred spatial direction in VSR, and can then be used to probe novel Lorentz-violating signatures. Taking \emph{q}\textsc{Bounce} sensitivity as a benchmark, we derive a first rough constraint for the neutron VSR parameter. Finally, we suggest alternative spin-flipping setups to better probe VSR effects and foresee potential future research directions.
 
\end{abstract}


\maketitle


\section{Introduction}

Lorentz symmetry certainly represents one of the fundamental pillars underpinning modern theoretical physics. In recent decades, however, the quest to explain some of the most puzzling enigmas in our understanding of nature has led to the development of entire research lines focusing on Lorentz-violating (LV) models. The latter were often embedded within more fundamental paradigms, such as string theory and quantum gravity\cite{collins2004lorentz,PhysRevD.39.683,PhysRevD.69.105009,mavromatos2007lorentz,carroll2001noncommutative}, from which the violation of Lorentz symmetry emerged at low energies as an effective feature.  \\
In this context, back in 2006, Cohen and Glashow presented a new mechanism to produce neutrino masses through Lorentz violation, which they called “very special relativity” (VSR) \cite{vsr1,vsr2}. Since then, the implications of the VSR landscape have been extensively studied in different areas; starting from the theoretical ones, such as the massiveness of photons \cite{PhysRevD.100.055029} and gravitons \cite{grav3,Santoni:2023uko,Bonilla:2025mrt}, to the more phenomenological ones, like corrections to scattering processes and anomalies \cite{Bufalo:2019kea,PhysRevD.103.075011,Bufalo:2020znk}, to the gyromagnetic factor of fermions~\cite{Koch:2022jcd} or to finite-temperature computations~\cite{PhysRevD.100.065024}. Within all these physical settings, calculations have been performed from the perspective of inertial systems, automatically excluding the possibility of accelerating observers or the effects coming from external gravitational backgrounds. Instead, these scenarios have been extensively explored using other LV approaches, especially within the Standard Model Extension (SME) \cite{Colladay:1998fq,PhysRevD.63.065008,lehnert2004dirac}. In fact, constraints on several SME parameters have already been obtained from tabletop experiments involving gravity, such as neutron interferometry observations and spectroscopy \cite{Kostelecky:2021tdf,kostelecky2011data,Ivanov:2019ouz,Ivanov:2021bvk}
However, the nature of the SME is fundamentally different from that of VSR in the sense that it only accounts for local corrections in the field Lagrangians. Indeed, VSR contributions are inherently non-local and, for that reason, they often transcend the boundaries of SME formulations. \\
Therefore, the main objective of this paper is to partially bridge the phenomenological gap between VSR and the domain of tabletop gravitational experiments. In particular, we focus on gravitational spectroscopy with ultracold neutrons (UCNs)~\cite{ignatovich1986physics}. This technique can generally reach very high sensitivities and is used in experiments like \emph{q}\textsc{Bounce} \cite{Abele:2008zz,abele2011qbounce,jenke2009q,Abele:2012dn} to investigate the quantized energy levels induced by the Earth's gravitational field when trapping UCNs with a reflecting floor. Under these circumstances, the simplest theoretical framework used to derive observationally-consistent predictions is based on the Schrödinger equation combined with a linear Newtonian potential
\begin{equation}
    i \hbar \partial_t \varphi = H_{qB} \,\varphi \,,
\end{equation}
where the Hamiltonian is 
\begin{equation} \label{HgFWlicase}
   {H}_{qB} =   m c^2 - m \vec g \cdot \vec x - \frac{\vec \partial^{\,2}}{2m}  \,.
\end{equation}
The gravitational acceleration $\vec g$ is typically assumed to be constant, homogeneous, and directed along the negative $\vec{\hat z} -$direction; $\vec g = - g \,\vec{\hat z}$ with $g>0$. The presence of the bottom mirror reflecting the freely falling neutrons is then simulated by setting an appropriate boundary condition for $\varphi$ at the mirror's height
\begin{equation} \label{boundaryqbounce}
    \varphi \left (z=0 \right )=0 \,.
\end{equation}
The solution of the time-independent Sch\"odinger equation featuring the Hamiltonian \eqref{HgFWlicase}, i.e.
\begin{equation}
    H_{qB} \, \varphi = E \, \varphi \,,
\end{equation}
is well-known and is given by Airy functions 
\begin{equation} \label{varphiairy}
    \varphi_N (z) = C_N \, Ai \left (\frac{z-z_N}{z_0} \right ) ,
\end{equation}
where we defined the quantities \cite{pitschmann2019schr}
\begin{equation} \label{airyparamdef}
    C_N = \frac{z_0^{-\frac12}}{Ai'\left(-\frac{z_N}{z_0}\right )} \, ,\,\,z_0 = \left(\frac{\hbar^2}{2 m^2 g} \right)^{\frac13},\,\, z_N = \frac{E_N}{m g }\,.
\end{equation}
Here, $Ai'(\zeta)$ represents the derivative of $Ai(\zeta)$ with respect to its argument $\zeta \equiv \frac{z-z_N}{z_0}$. The energy eigenvalues $E_N $ are determined by computing the zeros of the Airy function appearing in the quantization condition 
\begin{equation}
    Ai \left (-\frac{E_N}{m g z_0} \right )=0 \,,
\end{equation}
which is derived by enforcing the boundary constraint \eqref{boundaryqbounce} for the neutron eigenfunctions \eqref{varphiairy}.\\
Our goal is therefore to derive VSR corrections to the above picture starting from first principles, specifically, from the VSR equations of motion for a fermion in an accelerating (non-inertial) frame. \\
The outlook of the paper is as follows. After an introduction to the basic concepts of VSR and its Hamiltonian formulation, in Section \ref{chap2} we present the convention adopted for treating fermionic fields in curvilinear coordinates and describe the geometrical landscape in which we will work. Sec. \ref{chap3} is dedicated to the manipulation of the VSR equation in curvilinear coordinates, in order to derive its corresponding Hamiltonian. The non-relativistic (NR) limit is, instead, investigated and explicitly obtained in Sec. \ref{chap4}, where the machinery of the Foldy-Wouthuysen transformation is applied. Finally, in Sec. \ref{expconn} we connect our results to realistic experimental setups, while in Sec. \ref{chap5} we draw the respective conclusions and foresee potential future research directions. Some of the relevant calculations and details are provided in the Appendices at the end of the work and cited when needed.

\subsection{VSR Dirac Equation in Inertial Frames} \label{introvsrdirac}

Before we move on to the technicalities of treating spinors in curvilinear coordinates, let us introduce some basic concepts of VSR in inertial frames. We will refer to its $SIM(2)$ formulation \cite{vsr1}, implying no invariant tensors other than the usual Minkowski metric $\eta_{\mu\nu}$. Still, we shall assume the existence of a preferred null spacetime direction, labeled by $n^\mu$, which remains unaltered under $SIM(2)$ transformations
\begin{equation}
n^\mu \underset{SIM(2)}{ \Longrightarrow} \Omega\; n^\mu \,, \,\, \text{with }\, \Omega \in \mathbb{R}\,.
\end{equation}
This implies that quotients of scalar products involving the vector $n^\mu$ in both the numerator and the denominator are invariant under $SIM(2)$, since the rescaling parameter cancels out in the ratio. That allows us to write new non-local terms in the particle Lagrangians which break Lorentz symmetry down to the $SIM(2)$ subgroup. \\
Given this background, the modified Dirac equation in VSR can be written as
\begin{equation} \label{flatvsrdirac}
\left  ( i \slashed \partial - m + i \lambda \frac{\slashed n}{ n\cdot \partial}  \right ) \psi = 0 \,,
\end{equation}
where we use the ``slashed''-notation to represent contractions with gamma matrices. Note that $(n\cdot \partial)^{-1}$ should be regarded as an inverse operator, which can be expressed, for example, in the integral representation
\begin{equation}
    \frac{1}{n\cdot \partial } \psi  = \int_0 ^\infty d\alpha  \, e^{-\alpha n \cdot \partial} \psi \,.
\end{equation}
Although the VSR term may seem daunting at first sight, in the absence of interactions its only effect is to shift the physical mass of the fermion to the new value
\begin{equation}\label{vsrmassf}
    m_f = \sqrt{m^2 +2\lambda} \,.
\end{equation}
To avoid the encounter of runaway modes and remain consistent Appendix \ref{app:posdef}, we require the argument of the square root to be positive, so that throughout the paper we take 
\begin{equation}
    \lambda > - m^2 /2 \,.
\end{equation}
Moreover, it is customary and reasonable to think of $\lambda$ as a perturbative parameter due to the LV nature of VSR. For this reason, approaching the non-relativistic limit in Section \ref{chap4}, we will further assume $|\lambda| \ll m^2$.

\subsection{Hamiltonian Formulation of Inertial VSR} \label{hamform}

The time non-localities introduced by the new VSR operator in \eqref{flatvsrdirac} make it difficult to directly identify a Hamiltonian for the fermion, as is usually done for Lorentz-invariant (LI) systems. To overcome these complications, we have developed an ad hoc approach which generally involves three steps \cite{Santoni:2024coa}:
\begin{enumerate}
    \item Multiply the equations of motion (EOM) for $\psi$ by $-i n \cdot \partial$ from the left.
    \item Use the “squared” version of the EOM to replace the second-order time derivatives of $\psi$.
    \item Collect on one side terms that are linear in the time derivative of $\psi$ and write the EOM in the form of a Schrödinger equation with the appropriate Hamiltonian, i.e., $ i \partial_0 \psi = \mathcal H \psi$.
\end{enumerate}
Starting from \eqref{flatvsrdirac}, the first of the above steps leads us to the equation
\begin{equation} \label{towardvsrham}
( \,n\cdot \partial \, \slashed \partial +i m \, n \cdot \partial +\lambda \slashed n \,) \,\psi = 0 \,.
\end{equation}
Moving onto the second step, we have that the “squaring” procedure in the non-interacting case returns a Klein-Gordon (KG) equation for $\psi$ with an effective mass $m_f$ 
\begin{equation} \label{secordtd}
    (\partial^2 + m_f^2) \psi = 0  \,\,\to \,\, \partial_0^2 \psi = -(\partial_i \partial^ i + m_f^2) \psi \,.
\end{equation}
Now, we replace this expression for the second-order time derivative into \eqref{towardvsrham}. As for the third step, by isolating all terms proportional to first-order time derivatives on the left side, we end up with 
\begin{eqnarray} \label{flatsch1}
   && \left (1 - \frac{i}{m} (\gamma^0 n^ i +\gamma^ i) \partial_ i \right ) \, i \partial_ 0 \psi = \\
   && =
    \left ( \frac{1}{m} \gamma^0 ( \partial_i \partial^i +m^2_f) - \frac{1}{m } \gamma^ j n^i \partial_ i \partial_ j - i \, n^ i \partial_i -\frac{\lambda}{m} \slashed n \right) \psi \,. \nonumber
\end{eqnarray}
The operator on the left, which we define as
\begin{equation}
    \mathcal{B}^{-1}_\partial \equiv 1 - \frac{i}{m} (\gamma^0 n^ i +\gamma^ i) \partial_ i \,,
\end{equation}
satisfies the relation
\begin{eqnarray} \label{srlimit1}
    && m \gamma^0 - i \gamma^0 \gamma^ i \partial_i = \\
    && =\mathcal B_\partial \left (m \gamma^0+\frac{1}{m} \gamma^0 \partial_i \partial^i- \frac{1}{m } \gamma^ j n^i \partial_ i \partial_ j - i \, n^ i \partial_i \right) . \nonumber
\end{eqnarray}
Thus, inverting $\mathcal B_\partial$ on the right side of equation \eqref{flatsch1}, we arrive to a time-evolution equation from which we identify the following Hamiltonian 
\begin{equation}
    \mathcal{H}_{VSR} = m \gamma^0 -i \gamma^0 \gamma^i \partial_i + \frac{\lambda}{m} \mathcal B_\partial (\gamma^0 - \gamma^i n_i) \,.
\end{equation}
Clearly, the first two terms are equivalent to those in the Dirac Hamiltonian, while the third one represents the VSR contribution. Therefore, in the limit $\lambda\rightarrow 0$, we recover the LI scenario, as expected.\\
Dividing and multiplying the operator $\mathcal B_\partial$ on the right by $ (1 + \frac{i}{m} (\gamma^0 n^ i +\gamma^ i) \partial_ i  )$, we can re-express it as
\begin{equation}\label{nonlocalrelinertial}
    \mathcal B_\partial = \frac{1 + \frac{i}{m} \gamma^0 n^ i \partial_ i + \frac{i}{m}\gamma^ i \partial_ i }{1+\frac{1}{m^2} (n^i n^j - \delta^{ij}) \partial_ i \partial_j} \,.
\end{equation}
Hence, defining $\sigma^{ij } \equiv \frac12 [\gamma^i ,\gamma^j]$ and the non-local operator 
\begin{equation}\label{rinvdelta}
    \mathcal R_\partial^{-1} \equiv 1+\frac{1}{m^2} (n^i n^j - \delta^{ij}) \partial_ i \partial_j \,,
\end{equation}
the final form of the VSR Hamiltonian in inertial frames is given by
\begin{eqnarray}
    \mathcal{H}_{VSR} &=& (m + \frac{\lambda \,\mathcal R_\partial}{m}) \gamma^0 -i (1+ \frac{\lambda\,\mathcal R_\partial}{m^2})\gamma^0 \gamma^i \partial_i \\
    && - \frac{\lambda \, \mathcal R_\partial}{m}( \gamma^i n_i + \frac{i}{m} \gamma^0 \gamma^j n^ i n_j \partial_ i  + \frac{i}{m} \sigma^{ij} n_j \partial_ i ) \,. \nonumber
\end{eqnarray}
The momentum spectrum of the above Hamiltonian can be computed to be identical to the one of a free particle of mass $m_f$ in special relativity
\begin{equation} \label{vsrspectrumreal}
    E = \sqrt{m^2_f + \vec p^2} \,,
\end{equation}
in agreement with the spectrum anticipated by the KG-like dispersion relation in \eqref{secordtd}. Note that, even though $\mathcal H_{VSR}$ may not seem to satisfy the traditional Hermiticity criteria, its spectrum \eqref{vsrspectrumreal} is still strictly real, suggesting that the VSR Hamiltonian formulation is well-posed and unitarity can be preserved. Indeed, there exists an extensive literature on the possibility of meaningfully working with non-Hermitian Hamiltonians in both quantum mechanics (QM) and quantum field theory \cite{Bender:2007nj,alexandre2015foldy,Bender:2023cem,Bender:2020gbh}. In this context, the seeming lack of unitarity usually manifests when trying to naively interpret the behavior of certain open quantum systems with the toolkit of Hermitian QM (see, for example, the Lee model \cite{PhysRevD.71.025014,Giacosa:2020tha,Bender:2007nj}). This also points to the possibility that VSR corrections effectively characterize specific classes of open systems, describing the presence of complex interactions with some background entities or the environment \cite{Ilderton:2016rqk}.
\section{Theoretical Framework for Curvilinear Coordinates} \label{chap2}

When dealing with fermions in curvilinear coordinates, there are many conventions that need to be specified and clarified. First, we introduce a set of local observers by defining an orthonormal frame field $\{ \theta_{(a)} (x)\}$, which is connected to the coordinate basis $\{ \partial_{\mu} \}$ via the inverse tetrad $E_a^{\;\;\mu}(x)$ as follows
\begin{equation}
    \theta_{(a)}(x) = E_a^{\;\;\mu} (x)\, \partial_\mu  \,.
\end{equation}
The matrix $E_a^{\;\;\mu}$ is the inverse of the tetrad matrix $e^a_{\;\mu}$, which is in turn defined by its relation to the spacetime metric $g_{\mu\nu}$
\begin{equation}
    g_{\mu\nu} = e^a_{\;\mu} e^b_{\;\nu} \,\eta_{ab} \,,
\end{equation}
with the Minkowski tensor $\eta_{ab}$ representing the flat metric for the tangent spaces. Throughout the rest of the paper, we denote spacetime indices with Greek letters and tangent-space indices with Latin ones, as usual. Starting from the “flat” gamma matrices $\gamma^a$, we can define their “curved” version
\begin{equation}
    \underline \gamma^\mu \equiv E_{a}^{\;\;\mu} \, \gamma^{ a} ,
\end{equation}
which consistently satisfies the spacetime Clifford algebra
\begin{equation}
    \{ \underline \gamma^\mu, \underline \gamma^\nu \}= 2 \, g^{\mu\nu}  .
\end{equation}
Additionally, we recall the definition of the antisymmetrized product $\sigma^{ab}$ of two flat $\gamma-$matrices and introduce the block-diagonal four-by-four matrices $\vec \Sigma$ as
\begin{equation}
     \sigma^{ a b} \equiv \frac{1}{2} [\gamma^{ a} , \gamma^ { b}] \,\,\, ,\,\,\,
     \Sigma^i \equiv \left( \begin{array}{cc} \;  \sigma^i & 0 \\ 0 & \;\sigma^i \end{array}\right) ,
\end{equation}
with $\sigma^i $ being the two-dimensional Pauli matrices.
When separately taking into account spatial and time components for both spacetime and tangent-space indices, we will prevent any ambiguity with the tetrad indices (which mix both typologies) by always placing Latin indices first when reading from left to right. For other objects, any possible confusion is avoided by considering their original geometric nature. For brevity, Dirac indices will typically be omitted.\\
Finally, since the inverse power of $c$ is a natural parameter for performing formal expansions in the gravitational context, we shall retain the explicit $c$-units in our equations when needed.

\subsection{Spinor Connection in Curvilinear Coordinates}

To write down the Dirac equation in curvilinear coordinates \cite{de1962representations, Pollock:2010zz}, it is crucial to define the action of the spacetime covariant derivative $\mathcal D_\mu$ on a Dirac spinor $\psi$
\begin{equation}
    \mathcal D_\mu \psi = \partial_\mu \psi + \Gamma_\mu \psi \,.
\end{equation}
Here, the spinor connection $\Gamma_\mu$ is defined by the expression \cite{weinberg1972gravitation}
\begin{equation} \label{spinorconn}
    \Gamma_\mu \equiv \frac14 \sigma^{ a  b} \, \omega_{\mu\,  ab } = \frac14 \sigma^{ a  b} \, g_{\nu\rho} \, E_{ a}^{\;\;\nu} (\partial_\mu E_{ b}^{\;\;\rho} +\{^{\;\, \rho}_{ \mu\, \alpha}\} \, E_{ b}^{\;\;\alpha}) \,,
\end{equation}
where $\omega_{\mu\,  ab }$ is the Spin connection and $\{^{\;\, \rho}_{ \mu\, \alpha}\}$ are the Christoffel symbols defined in General Relativity (GR)
\begin{equation}
    \{^{\;\, \rho}_{ \mu \,\, \alpha}\} = \frac{1}{2} g^{\rho \beta} ( \partial_\mu g_{\alpha \beta} +\partial_\alpha g_{\mu \beta} -\partial_\beta g_{\mu \alpha})\,.
\end{equation}
We emphasize that both the curved gamma matrices and the tetrads are designed to be automatically covariantly constant \cite{Huang:2005rn, parker1980one,huang2009hermiticity}
\begin{eqnarray}
    \mathcal D_\mu \underline \gamma^\nu &=&  \partial_\mu \underline \gamma^\nu + \{^{\;\, \nu}_{ \mu \,\, \rho}\} \underline \gamma^\rho - [ \Gamma_\mu, \underline \gamma^\nu ] = 0 \,, \\
    \mathcal D_\mu e^a_{\;\nu} &=& \partial_\mu e^a_{\;\nu} - \{^{\;\, \rho}_{ \mu \,\, \nu}\} \, e^a_{\;\rho} + {{\omega_{\mu }}^a}_b  e^b_{\;\nu} =0 \,,
\end{eqnarray}
which is something to keep in mind during the upcoming calculations.

\subsection{Spacetime for Accelerated Frames}

According to the GR perspective, a laboratory on Earth's surface is constantly accelerating outward with a certain acceleration $\vec a$. This is because, instead of following a geodesic, it remains at a fixed distance from the center of the Earth. From the laboratory point of view, that leads to an “apparent” acceleration $\vec g=-\vec a$ for freely falling objects, which is naturally ascribed to a gravitational force, as in the familiar Newtonian picture. \\ 
As long as we want to describe sufficiently small patches of spacetime, meaning small spatial volumes for small time periods \cite{Misner:1973prb}, and therefore deal with (almost) uniform gravitational fields, as happens in the context of spectroscopic experiments with UCNs \cite{Kostelecky:2021tdf}, the laboratory acceleration $\vec a$ can be safely considered constant and homogeneous. Thus, a natural way to geometrically describe this accelerated setting is through the following spacetime line element $ds^2$ and metric tensor $g_{\mu\nu}$
\begin{eqnarray} \label{rindlerelement}
    && ds^2 = \left (1+ \frac{\vec a \cdot \vec x}{c^2} \right )^2 d (ct)^2 - d\vec x ^{\,2} \,, \\
    && g_{\mu\nu} = diag \left \{ \left (1+ \frac{\vec a \cdot \vec x}{c^2} \right)^2 ,-1,-1,-1 \right \}, \nonumber
\end{eqnarray}
where we used $x^0 = ct$, $\vec a \cdot \vec x = a^ i x^ i  $ and momentarily restored $c-$units. For convenience, in the rest of the article we take the $z-$axis to coincide with the acceleration direction $\vec a = a \, \vec{\hat z} = -\vec g $. \\
The $ds^2$ in \eqref{rindlerelement} agrees with the one obtained by charting flat spacetime with Kottler-Møller or Rindler coordinates \cite{Rindler1969-RINER,moller1943homogeneous}, which are the natural set of coordinates adopted by (non-inertial) observers undergoing uniformly accelerated motion. Hence, we should stress that the Riemann tensor associated with \eqref{rindlerelement} is zero
\begin{equation} \label{zeroriemann}{{R_{\mu\nu}}^\alpha_{\;\;\beta} } = \partial_\mu \{^{\;\, \alpha}_{ \nu \,\, \beta}\} - \partial_\nu \{^{\;\, \alpha }_{ \mu \,\, \beta}\}+\{^{\;\, \alpha }_{ \mu \,\, \rho}\}\{^{\;\, \rho }_{ \nu \,\, \beta}\} - \{^{\;\, \alpha }_{ \nu \,\, \rho}\}\{^{\;\, \rho }_{ \mu \,\, \beta}\} = 0 \,,
\end{equation}
meaning that tidal effects $\propto |x|^2$ are not depicted in this scheme. However, for the level of precision required in the present work, they would be irrelevant to the final result anyway. Neglecting $|x|^2$ contributions, the geometric structure above is also approximately equivalent to the spacetime element measured by a static observer resting at some fixed distance $R$ from the center of a Schwarzschild background. In fact, the latter represents a non-rotating and accelerating observer whose proper reference frame could be described through Fermi-Walker coordinates \cite{maluf2008construction,klein2008general}. \\
The tetrads that naturally describe an observer standing still in the laboratory, i.e. one that is stationary relative to the accelerated coordinate chart $\{x^\mu\}$ characterized by \eqref{rindlerelement}, are
\begin{eqnarray} \label{rindlertetrads}
   && e^a_{\;\;\mu} = diag\{V,1,1,1\} \,\, \text{with }\,\, V \equiv 1+\frac{\vec a \cdot \vec x }{c^2} \,, \nonumber\\
   && E_a^{\;\;\mu} = diag\{\frac{1}{V},1,1,1\} \, ,
\end{eqnarray}
leading to the curved gamma matrices below
\begin{eqnarray}
    \underline \gamma^0 =  E^{\;\;0}_a \gamma^a = \frac{1}{V} \gamma^0 \,\,,\,\,\,
    \underline \gamma^i = E^{\;\;i}_a \gamma^a = \gamma^i\,. 
\end{eqnarray}
In Appendix~\ref{app:rindlerquantities}, we include all other geometric quantities relevant to our calculations, such as the components of the spinor connection $\Gamma_\mu$ and the Christoffel symbols.

\section{VSR Dirac Equation in Curvilinear Coordinates} \label{chap3}

Following the usual prescription of General Relativity, the extension of equation \eqref{flatvsrdirac} to the context of curvilinear coordinates is obtained through the replacement of partial derivatives $\partial_\mu$ with covariant derivatives $\mathcal D_\mu$ 
\begin{equation} \label{vsreomgrav}
    (i \slashed {\mathcal D} -m +i \lambda \frac{\slashed n}{n \cdot {\mathcal D}}) \psi = 0 \,,
\end{equation}
where the slashed notation is here used to indicate a contraction with the curved gamma matrices $\underline \gamma^\mu$. The constancy condition of $n^\mu$ in inertial frames should also be adapted to curvilinear coordinates. The simplest way to do it is to assume its covariant derivative to vanish
\begin{equation} \label{devcovn}
    \mathcal D_ \mu n^\nu = 0 \,.
\end{equation}
Although it may not be the unique way to generalize the concept of a constant $n^\mu$, in this paper we will use (\ref{devcovn}) as a working hypothesis. In fact, because of the zero curvature \eqref{zeroriemann}, we do not encounter no-go constraints that would forbid the existence of such a $n^\mu$ \cite{PhysRevD.91.065034,Kostelecky:2021tdf}, and we can always find solutions to the above equation. This also means that $n^\mu$ would now be position-dependent in general, while still being lightlike. \\
Alternatively, we might take $n \cdot \mathcal D n^\mu = 0$, the solution of which is a set of null geodesics. Both possibilities allow us to define the VSR equation \eqref{vsreomgrav} without any ambiguity, since they eliminate ordering problems with the numerator and the denominator of the non-local term. The preference for one over the other depends solely on the assumed fundamental origin of the preferred spacetime direction in VSR.

\subsection{“Curved” Hamiltonian in VSR}

The procedure to obtain the Hamiltonian associated with equation~\eqref{vsreomgrav} is completely analogous to the one followed for the inertial case in Section \ref{hamform}. Indeed, since the curvature tensor vanishes, the commutator of covariant derivatives acting on $\psi $ vanishes accordingly
\begin{equation}
      [{\mathcal D}_\mu, {\mathcal D}_\nu]\psi = -\frac{1}{4} R_{\mu\nu\rho\sigma} \underline\gamma^\rho \underline \gamma^\sigma \psi = 0\,.
\end{equation}
From that, it is straightforward to show that the “squared” EOM will have once again the KG structure, but now in curvilinear coordinates. Recalling the definition \eqref{vsrmassf} of $m_f$, we find
\begin{equation}
    (g^{\mu\nu} {\mathcal D}_\mu {\mathcal D}_\nu + m^2_f)\psi =0 \,,
\end{equation}
implying the relation
\begin{equation} \label{kgrel2}
    {\mathcal D}_0^2 \psi = - (g^{00})^{-1}( {\mathcal D}_i {\mathcal D}^i +m^2_f ) \psi  = -V^2 ( {\mathcal D}_i {\mathcal D}^i +m^2_f ) \psi\,.
\end{equation}
At this point, to follow steps 1 and 2 from Section \ref{hamform}, we need to first expand the contractions appearing in the curvilinear equivalent of equation \eqref{towardvsrham}, i.e.
\begin{equation} \label{vsreomgrav2}
    ( n\cdot {\mathcal D} \slashed {\mathcal D} +i m n\cdot {\mathcal D} + \lambda \slashed n) \psi = 0 \,.
\end{equation}
That gives the two following identities:
\begin{eqnarray} \label{calculationnDDgrav}
    n\cdot {\mathcal D} \slashed {\mathcal D} &=& n^0 \underline \gamma^0 {\mathcal D}_0 ^2 + (\underline \gamma^0 n^ i +n^0\underline \gamma^ i) {\mathcal D}_ i {\mathcal D}_0 + \underline \gamma^ j n^i {\mathcal D}_ i {\mathcal D}_ j \nonumber \\
    &=& \frac{\,n^0 }{V}  \gamma^0 {\mathcal D}_0 ^2 + ( \frac{\,n^ i }{V} \gamma^0 +n^0  \gamma^ i) {\mathcal D}_ i {\mathcal D}_0 +  \gamma^ j n^i {\mathcal D}_ i {\mathcal D}_ j \,, \nonumber \\
    im n\cdot {\mathcal D} &=& im \, n^0 {\mathcal D}_0 + im \, n^ i {\mathcal D}_i \, .
\end{eqnarray}
Replacing these expressions in \eqref{vsreomgrav2} together with \eqref{kgrel2} and factorizing the time covariant derivative of $\psi$ in the left-hand side, we obtain
\begin{eqnarray} \label{almostschemgrav2.8}
    &i& \left (1 - \frac{i}{m} (\gamma^0 \frac{ n^i}{V n^0 } + \gamma^i ) \mathcal D_i \right ) {\mathcal D}_0 \psi = \\
    &&= \left ( m V \gamma^ 0 + \frac{V}{m} \gamma^0 {\mathcal D}_i {\mathcal D}^i - i  \frac{n^i}{n^0} {\mathcal D}_i - \frac{1}{m}\frac{n^i}{n^0}  \gamma^j {\mathcal D}_i {\mathcal D}_j \right. \nonumber \\
    && \;\;\;\;\;\;\;\; \left . + \frac{\lambda}{m} V \gamma^ 0 - \frac{\lambda}{m} \frac{n_i}{n^0} \gamma^i \right) \psi \,.\nonumber
\end{eqnarray}
Evaluating the covariant derivatives along spatial directions we observe that, since $\Gamma_i=0$ (see Appendix~\ref{app:rindlerquantities}), they always reduce to ordinary partial derivatives 
\begin{eqnarray}
    {\mathcal D}_i \psi &=& \partial_i \psi + \Gamma_i \psi = \partial_i \psi\,, \\
    {\mathcal D}_j {\mathcal D}_i \psi &=& \partial_j {\mathcal D}_i \psi + \Gamma_j {\mathcal D}_i\psi -\{^{\;\, \mu }_{ j \,\,\, i} \} {\mathcal D}_\mu \psi =\partial_j \partial_i \psi  \,,\nonumber
\end{eqnarray}
except when time components are involved, e.g.
\begin{equation}
    \mathcal D_i \mathcal D_0 \psi = (\partial_i -\{^{\;\, 0 }_{ i \,\,\, 0} \} + \Gamma_i  ) \mathcal D_0 \psi = (\partial_i -\frac{1}{V} \partial_i V   ) \mathcal D_0 \psi 
    \,.
\end{equation} 
Taking that into account, we can write equation \eqref{almostschemgrav2.8} as 
\begin{eqnarray} \label{almostschemgrav3}
    i {\mathcal D}_0 \psi =  \mathcal B_{\mathcal D} && \left ( m V \gamma^ 0 + \frac{V}{m} \gamma^0 {\partial}_i { \partial}^i - i  \frac{n^i}{n^0} {\partial}_i - \frac{n^i \gamma^j}{m \,n^0}  {\partial}_i {\partial}_j \right. \nonumber \\
    && \;\;\;\; \left . + \frac{\lambda}{m} V \gamma^ 0 - \frac{\lambda}{m} \frac{n_i}{n^0} \gamma^i \right) \psi \,,
\end{eqnarray}
where we defined the operator
\begin{equation}
    \mathcal B ^{-1}_{{\mathcal D}} \equiv 1 - \frac{i}{m} (\gamma^0 \frac{ n^i}{V n^0 } + \gamma^i ) (\partial_i - \frac{1}{V} \partial_i V )  \, .
\end{equation}
The latter satisfies a relation analogous to \eqref{srlimit1} 
\begin{eqnarray}\label{lambdazerocheck}
    && m V \gamma^0 - i V \gamma^0 \gamma^i \partial_i = \\
    && =\mathcal B_{\mathcal D} \left( m V \gamma^ 0 + \frac{V}{m} \gamma^0 \partial_i \partial^i - i  \frac{n^i}{n^0} \partial_i - \frac{1}{m} \frac{n^i}{n^0} \gamma^j \partial_i \partial_j \right ) , \nonumber
\end{eqnarray}
so that we recover the correct “curved” Dirac equation when $\lambda \to 0$. Bearing that in mind, while also recalling 
\begin{equation}
    \mathcal D_0 \psi = \partial_0 \psi +\Gamma_0 \psi=\partial_0 \psi +\frac{1}{2} \gamma^0 \gamma^i \partial_i V \psi
\,,
\end{equation}
we can identify from \eqref{almostschemgrav3} and \eqref{lambdazerocheck} the following simplified expression for the Hamiltonian 
\begin{equation} \label{hamiltoniangVSRtext}
    \mathcal H^g _{VSR}  = m V \gamma^0 - i \gamma^0 \gamma^i (V \partial_i +\frac12 \partial_i V) +\frac{\lambda \mathcal B_{\mathcal D}}{m}  V ( \gamma^0 - \tilde n_i \gamma^i ) \,,
\end{equation}
where we introduced the vector $\tilde n^ i$ as
\begin{equation}
    \tilde n ^ i \equiv \frac{n^ i }{V n^ 0} \,,
\end{equation}
which is independent of the space-position $\vec x$ (more details in Section \ref{nprop}). We can now eliminate the gamma structure in the denominator of $\mathcal B_{\mathcal D}$ by observing that
\begin{eqnarray}
    &&\mathcal B_{\mathcal D} V( \gamma^0 - \tilde n_i \gamma^i )= \\
    &&= \mathcal R_{\mathcal D} V\left ( \gamma^0 - \tilde n_i \gamma^i -\frac{i}{m} ( \gamma^0 \gamma^i +\gamma^0 \gamma^j \tilde n^i \tilde n_j +\sigma^{ij} \tilde n_j) \partial_i \right ) , \nonumber
\end{eqnarray}
with the new non-local operator $\mathcal R_{\mathcal D}$ defined to be
\begin{equation}
    \mathcal R^{-1}_{\mathcal D} \equiv 1 +\frac{\tilde n^i \tilde n^j + \eta^{ij} }{m^2} \left (\partial_i\partial_j - \frac{2\,\partial_i V}{V}  \partial_j + \frac{2 \,\partial_i V \partial_j V}{V^2} \right ) .
\end{equation}
Thus, we can express Eq.~\eqref{hamiltoniangVSRtext} as
\begin{eqnarray} \label{Hgvsr1}
   \mathcal{H}^g_{VSR} &=& (1+\frac{\lambda \mathcal R_{\mathcal D} }{m^2}) V \gamma^0 ( m - i \gamma^i \partial_i ) - \frac{i}2 \gamma^0 \gamma^i \partial_i V \\
    && -\frac{\lambda}{m} \mathcal R_{\mathcal D} V \left ( \tilde n_i \gamma^i  +\frac{i}{m} (  \gamma^0 \gamma^j \tilde n^i \tilde n_j +\sigma^{ij} \tilde n_j) \partial_i \right ) . \nonumber
\end{eqnarray}
Finally, using the identity below
\begin{equation}
    \mathcal R_{\mathcal D} V  =  V \, \left (1 + \frac{1}{m^2}(\tilde n^i \tilde n^j + g^{ij}) \partial_i \partial_j \right )^{-1} =V  \tilde {\mathcal R}_\partial \,,
\end{equation}
we rewrite the non-inertial VSR Hamiltonian
\begin{eqnarray} \label{Hgvsr1.5}
   \mathcal{H}^g_{VSR} &=& V (1+\frac{\lambda \tilde {\mathcal R}_\partial }{m^2})  ( m \gamma^0 - i \gamma^0 \gamma^i \partial_i ) - \frac{i}2 \gamma^0 \gamma^i \partial_i V \nonumber\\
    && -\frac{\lambda V}{m} \tilde {\mathcal R}_\partial  \left ( \tilde n_i \gamma^i  +\frac{i}{m} (  \gamma^0 \gamma^j \tilde n^i \tilde n_j +\sigma^{ij} \tilde n_j) \partial_i \right ) , \nonumber\\
\end{eqnarray} 
with the tilde in $\tilde{\mathcal R} _\partial$ indicating the use of $\tilde n^i$ over $n^i$ in its definition~\eqref{rinvdelta}.

\subsection{Properties of the Vector $\tilde n^i$} \label{nprop}

Let us stop for a moment to ponder over the properties of the vector $\tilde n^i$. By considering the constancy condition \eqref{devcovn} for spatial components, we find
\begin{equation}
   0 = {\mathcal D}_i n^j = \partial_i n^j + \{^{\;\, j }_{ i \,\,\, \mu} \} n^\mu = \partial_i n^j \,,
\end{equation}
where we employed the results in Appendix~\ref{app:rindlerquantities}. The implementation of \eqref{devcovn} for the time component, instead, yields
\begin{equation}
    0 = {\mathcal D}_i n^0 = \partial_i n^0 + \{^{\;\, 0 }_{ i \,\,\, \mu} \} n^\mu  = \partial_i n^0 +\frac{1}{V} \partial_i V n^0 \,.
\end{equation}
Multiplying the last equation by $V$, we get
\begin{equation}
    V \partial_i n^0 +\partial_i V n^0 = \partial_i (V n^0) = 0\,.
\end{equation}
Thus, as already mentioned, the combination of the $n^\mu-$components as defined into $\tilde n^i$ is such that the latter does not depend on spatial coordinates
\begin{equation}
    \partial_ i \tilde n^j = \partial_i (\frac{n^j}{n^0 V}) = 0\,.
\end{equation}
Furthermore, since $n^\mu$ is lightlike
\begin{equation}
    g_{00} (n^0)^2 = V^2 (n^0)^2 = - n_i n^i = n^i n^i = |\vec n |^2 \,,
\end{equation}
the new $\tilde n ^i$ turns out to be a unit vector
\begin{equation}
    |\vec {\tilde n} |^2 = \frac{|\vec n|^2}{(n^0 V)^2} = 1\,.
\end{equation}
Hence, we can identify $\tilde n^i$ as a time-dependent unit vector, playing a similar role as $n^i$ in inertial frames, where the VSR vector is usually rescaled to the form $(1,\vec{\hat n})$. In addition, by considering the remaining two conditions that arise from Eq.~\eqref{devcovn}
\begin{eqnarray}
    && 0= {\mathcal D}_0 n^i = \partial_t n^i + V\partial_i V n^0\,,\\
    && 0= {\mathcal D}_0 n^0 = \partial_t n^0 + \frac{1}{V} \partial_i V n^i \,, \nonumber
\end{eqnarray}
we obtain the time-evolution equation for $\vec{\tilde n}$
\begin{eqnarray} \label{timeevofn}
    \partial_t \tilde n^i &=& \frac{1}{V} \partial_t (\frac{n^i}{n^0}) = \frac{1}{V n^0} \partial_t n^i - \frac{n^ i}{V (n^0)^2} \partial_t n^0 \nonumber \\
    &=& \tilde n^i \tilde n^k \partial_k V -\partial_i V \,\,\propto \,\, \nicefrac{1}{c} \,,
\end{eqnarray}
where we highlighted its $c-$scaling at the end. This behavior will be helpful for some considerations in the next sections. Clearly, solving equation \eqref{timeevofn} allows us to find the dynamics of the $\vec{\tilde n}-$components and to see how this special direction changes over time (from the laboratory perspective). \\
After some calculations, included in the Appendix~\ref{app:evoofn}, we find that $\vec{\tilde n}$ does not undergo any motion in the equatorial plane. However, its orientation does change in the azimuthal direction and tends to anti-align with the local acceleration $\vec a$. 

\subsection{Restoring Hermiticity and Unitarity}

Like in the inertial scenario, the non-inertial VSR Hamiltonian contains non-Hermitian terms, as shown in Appendix~\ref{app:spinmetric}. For solutions of the Schrödinger-like equation determined by \eqref{Hgvsr1.5}, those contributions spoil the conservation of probability amplitudes
\begin{equation}
    \partial_t(\psi,\psi)= \partial_t \int d^3x \, \psi^\dagger \psi \neq 0 \,.
\end{equation}
However, the latter can be restored by introducing an appropriate spinorial metric $ G$ into the definition of the scalar product $(\cdot\,,\cdot )_G$ as
\begin{equation}
    (\psi,\psi)_{G} \equiv \int d^3x \, \psi^\dagger \, G \,\psi \,,
\end{equation}
where $G$ has to satisfy \cite{Ju:2021vvs}
\begin{equation}\label{definingrelG}
    \partial_t G + i (\mathcal H ^g_{VSR})^\dagger  G - i G \, \mathcal H ^g_{VSR} =0 \, .
\end{equation}
This procedure is applied, for example, in the case of non-homogeneous gravitational fields \cite{Obukhov:2011ks}, where the spinorial metric gets modified to $\sqrt{-g} \,{E_0}^0$ (factor that equals unity in our case). For more details, see \cite{parker1980one,arminjon2006post,huang2009hermiticity} and references therein.\\
In the VSR scenario, we obtain the spinorial metric
\begin{equation} \label{Gexpr}
    G= 1+\frac{\lambda }{m^2} (1 - \gamma^0 \gamma^ i \tilde n_i ) \, \tilde {\mathcal R}_\partial \,,
\end{equation}
which is Hermitian and positive definite, as expected for a good metric. More details on its derivation can be found in Appendix~\ref{app:spinmetric}. Crucially, for every spinorial metric one can always find a (non-unique) matrix $\Delta$ such that it holds the decomposition
\begin{equation}
    G = \Delta^\dagger \Delta \,.
\end{equation}
The latter suggests a redefinition $\psi^\Delta = \Delta \psi$ of the spinor, which implies the transformation of the original non-Hermitian Hamiltonian $\mathcal H$ into a new one 
\begin{equation} \label{transfhamdelta}
    {\mathcal H}^\Delta = \Delta \mathcal H \Delta ^{-1} + i \Delta^{-1}\partial_t \Delta \,,
\end{equation}
that is Hermitian with respect to the standard inner product (i.e. using $G=1$) and describes the same dynamics. For our case, we obtain the following exact result
\begin{equation} \label{deltaexprexact}
    \Delta =
    \frac{1+\frac{\lambda \tilde {\mathcal R}_\partial}{m^2} + \sqrt{1+\frac{2\lambda \tilde {\mathcal R}_\partial}{m^2}}-\frac{\lambda \tilde {\mathcal R}_\partial}{m^2} \, \gamma^0 \gamma^i \tilde n_i}{2 ^\frac12\,  \sqrt{ 1+\frac{\lambda \tilde {\mathcal R}_\partial}{m^2} + \sqrt{1 + \frac{2\lambda \tilde {\mathcal R}_\partial}{m^2}}}  }  \,.
\end{equation}
In the hypothesis of VSR emerging as a minor departure from the LI setting, we can reasonably take $|\lambda|\ll m^2$, as already mentioned in Section \ref{introvsrdirac}. Therefore, expanding formula \eqref{deltaexprexact} for small $\lambda-$values, we find
\begin{equation}
    \Delta = 1 + \frac{\lambda }{2m^2} (1- \gamma^ 0 \gamma^i \tilde n_i)\, \tilde {\mathcal R}_\partial + O(\lambda^2) \,,
\end{equation}
and replacing it back in \eqref{transfhamdelta}, we obtain the final expression of the transformed Hamiltonian at first order in $\lambda$
\begin{widetext}
\begin{eqnarray} \label{Hgvsrhermitian}
    ( {\mathcal H}^g_{VSR})^\Delta &=& m V \gamma^ 0 + \frac{\lambda}{m} (V \tilde {\mathcal R}_\partial - \frac12 [ \tilde r_\partial, V ] \tilde{\mathcal R}^2_\partial)\, \gamma^0  -\frac{\lambda}{2m} [ \tilde r_\partial , V] \tilde{\mathcal R}^2_\partial \gamma^i \tilde n_i \nonumber \\
    && - i \gamma^0 \gamma^i (V \partial_i + \frac12 \partial_iV ) - i \frac{\lambda}{m^2} \gamma^0\gamma^i (V \tilde{\mathcal R}_\partial \partial_i + \frac12 \mathcal \partial_i V \tilde {\mathcal R}_\partial - \frac12 [ \tilde r_\partial, V ] \tilde{\mathcal R}^2_\partial \partial_i ) \\
    && -i \frac{\lambda}{m^2 }\gamma^0 \gamma^j \tilde n^i \tilde n_j ( V \tilde{\mathcal R}_\partial \partial_i + \frac12 \partial_i V \tilde {\mathcal R}_\partial - \frac12 [ \tilde r_\partial , V ] \tilde{\mathcal R}^2_\partial \partial_i)  + i \frac{\lambda}{2m^2} \sigma^{ij} \tilde n_j ( \partial_i V \tilde{\mathcal R}_\partial - [ \tilde r_\partial, V] \tilde{\mathcal R}^2_\partial \partial_i) \,,\nonumber
\end{eqnarray}
\end{widetext}
where we defined $ \tilde r_\partial $ to be 
\begin{equation}
     \tilde r_\partial \equiv \frac{1}{m^2} (\tilde n^i \tilde n^j + \eta^{ij}) \partial_i \partial_j \,,
\end{equation}
with $\eta^{ij}$ being the spatial components of the Minkowski metric. In the rest of the paper, for the sake of simplicity, we drop the $\Delta-$apex on the transformed Hamiltonian \eqref{Hgvsrhermitian} and spinors.

\section{Non-Relativistic Limit} \label{chap4}

In this section, we develop a systematic approach to obtain the NR limit of the Hamiltonian \eqref{Hgvsrhermitian}. Our analysis constitutes a natural adaptation of the Foldy–Wouthuysen (FW) transformation, which is commonly employed to access the NR regime in the form of a Schrödinger equation with systematic corrections in inverse powers of the Dirac mass \cite{foldy1950dirac,Bjorken:1965sts,Berestetskii:1982qgu}. However, as we shall see, the choice for the expansion parameter is more subtle in this particular instance.

\subsection{Foldy-Wouthuysen Transformation}

The standard way to proceed with the FW transformation is the following. The first step is to divide the Hamiltonian into the so-called “even” and “odd” contributions labeled, respectively, by $\mathcal E$ and $\Theta$
\begin{equation}
    \mathcal H = m\gamma^0 + \mathcal E + \Theta \,.
\end{equation}
These operators are identified according to the number of spatial gamma matrices involved and satisfy specific commutation relations
\begin{equation}
    [\mathcal E, \gamma^0] =0 \,\,,\,\,\,\, \{\Theta , \gamma^0 \} =0\,.
\end{equation}
Then, the FW unitary transformation $ U\equiv e^{i S}$ consists of a change in the spinor and the Hamiltonian of the type
\begin{equation} \label{fwtransfh}
   \psi'  = U \psi   \,\,,\,\,\,\, \mathcal H' = U \mathcal H U^\dagger - i\, U \partial_t U ^\dagger  \,.
\end{equation}
The operator $S$ in the exponent of $U$ is defined as
\begin{equation} \label{operS}
    S \equiv - \frac{i}{2m c^2} \gamma^0 \Theta \, .
\end{equation}
Working in the NR domain, where $  mc^2$ represents the largest energy scale of the system, we can perform a formal expansion of the transformed Hamiltonian \eqref{fwtransfh} up to any order in $\nicefrac{1}{m}$
\begin{eqnarray}\label{HFWcommut}
    \mathcal{H'} &=& e^{iS} \mathcal H e^{-iS} - i e^{iS} \frac{\partial }{\partial t} (e^{-iS}) \\
    &=& \mathcal H + i [S,\mathcal H] + \frac{i^2}{2!} [S,[S,\mathcal H]] +\frac{i^3}{3!} [S,[S,[S,\mathcal H]]] \nonumber \\
    && +\frac{i^4}{4!} [S,[S,[S,[S,\mathcal H]]] \nonumber \\
    && - \dot S - \frac{i}{2} [S,\dot S] + \frac16 [S,[S,\dot S]] + {\mathcal{O}}(m^{-4}) \nonumber\,.
\end{eqnarray}
In this way, the particle-antiparticle sectors of a Dirac system can be decoupled up to the desired precision by iteratively applying FW transformations. In fact, the operator \eqref{operS} is specially designed to cancel out the leading-order odd contributions in each FW iteration. 

\subsection{Parameters controlling the perturbative expansion and their order
} \label{sec:orderofpert}

Although the FW procedure is generally carried out using an expansion in $\nicefrac{1}{m}$ \cite{Bjorken:1965sts}, formal $\nicefrac{1}{c}-$expansions are better suited for gravitational (or non-inertial) contributions, because $c$ also plays a role in the derivation of the weak-field (or low-acceleration) limit. This is common practice in post-Newtonian (PN) approximations implemented in other gravitational contexts \cite{chandrasekhar1965post,nelson1990post}. The usual terminology used to classify different groups of terms in the expansion is shown in Table \ref{tab:PNcorresp}.
\begin{table}[ht]
\centering
\begin{tabular}[t]{ccc}
\toprule
$\;\;${$\nicefrac{1}{c}-$order}$\;\;$ & $\;\;$PN-equivalent$\;\;$ \\ 
    \midrule
    {$1$} & 0PN \\
    $c^{-1}$ & 0.5PN \\
    $c^{-2}$ & 1PN \\
    \bottomrule
\end{tabular} 
\caption[Post-Newtonian perturbative scheme]{Correspondence table between the $\nicefrac{1}{c}-$expansion and the post-Newtonian one (also look at Fig.1 in \cite{bern2019black}).}
\label{tab:PNcorresp}
\end{table}\\
However, the presence of VSR corrections, which often involve terms proportional to $c^2$, undermines this strategy. Hence, sticking to the parameter $m$ allows us to develop the following argument to identify leading order (LO) and next-to-leading order (NLO) perturbations: Let us consider the reduced Hamiltonian $\mathcal H /mc^2$ and recall that
\begin{equation}
    v \equiv V-1 = \frac{\vec a \cdot \vec x }{c^2} \, \propto \, \nicefrac{1}{c^2} \,.
\end{equation}
Then, there are only three dimensionless arrangements of parameters that we could expect to appear in the non-relativistic limit given by the FW procedure
\begin{equation} \label{pertparam}
   \left \{ \frac{\lambda}{m^2} , \frac{p}{mc} , v \right \} ,
\end{equation}
with $p$ representing momentum operators.

\subsubsection{Leading Order -- 0PN }

Starting from \eqref{pertparam}, the combinations we can construct at order $\nicefrac{1}{c^2}$ and up to first order in $\lambda$ are 
\begin{equation}\label{perturbativestructures}
    \nicefrac{1}{c^2} \,\,\, - \,\,\, \left \{ \left(\frac{p}{mc}\right)^2, \frac{\lambda}{m^2} \left(\frac{p}{mc}\right)^2, v, \frac{\lambda}{m^2} v \right \} , 
\end{equation}
where, in constructing the above terms, we considered that non-VSR terms (i.e. terms not proportional to $\lambda$) cannot have odd powers of $p$, because there is no way to form scalars with an odd number of vectors. Furthermore, we point out that new non-trivial VSR structures (featuring $\tilde n^ i$) should appear with $v$, ensuring that they vanish in the zero-acceleration limit, thus recovering the correct inertial scenario in VSR.\\
As we can see from \eqref{perturbativestructures}, all $\nicefrac{1}{c^2}$-corrections to ${\mathcal H}/{mc^2}$ are at most of order $\nicefrac{1}{m^4}$. Then, it is sufficient to perform the FW transformation up to order $\nicefrac{1}{m^3}$, to obtain all 0PN Hamiltonian contributions. Clearly, during this process, 0.5PN and 1PN corrections, such as terms $\propto v^2$, may also be present, but they are easy to detect and discard safely. In fact, any term $A$ that becomes negligible at the $n$th FW step cannot yield non-negligible effects in subsequent steps. That is because all its contributions to the $(n+1)-$iteration are in some way derived from products of the type
\begin{equation}
    A  \,\frac{\mathcal E^{(n)}}{mc^2} \,\,,\,\,\,  A \, \frac{\Theta^{(n)}}{mc^2} \text{ or  higher order} \,,
\end{equation}
which cannot produce lower PN$-$contributions, since the even and odd operators are at most $\propto c^2$ even for the first iteration, as we shall see later. An equivalent reasoning is valid for terms that are already higher order in $\nicefrac{1}{m}$.

\subsubsection{Next-to-Leading Order -- 0.5PN } \label{NLOsec}

In going further into the $\nicefrac{1}{c}-$expansion, new structures are expected to appear. 1PN corrections are already extremely tiny in the standard LI context \cite{PhysRevD.109.064085}, so we focus only on 0.5PN. Moreover, in the non-rotating LI case, there are no 0.5PN components, implying that in our calculations we can safely expect all of them to be VSR related. Repeating the same analysis as before for $\nicefrac{1}{c^3}-$combinations, we obtain just one viable option
\begin{eqnarray}\label{perturbativestructures2}
    && \nicefrac{1}{c^3} \,\,\, - \,\,\, \left \{ \frac{\lambda}{m^2} \frac{p}{mc} v \right \} \,. 
\end{eqnarray}
Interestingly, the latter is $\propto \nicefrac{1}{m^3}$ in the reduced Hamiltonian $\mathcal H / mc^2$. Thus, in order to find all 0.5PN contributions, it is sufficient to perform the FW expansion up to $\nicefrac{1}{m^2}$. The complete 0.5PN Hamiltonian is then obtained adjoining these new perturbations with the previous LO result. Note that, due to the relatively low order of our approximations, it will not be necessary to worry about the eventual discrepancies emerging between exact and iterative approaches to the NR limit \cite{Silenko:2024zkt,neznamov2009foldy,Silenko:2009if}.

\subsection{Non-Relativistic Hamiltonian at 0PN}

At this point, we can safely proceed with the implementation of the FW procedure. As outlined above, we start with the LO evaluation at 0PN. For convenience, we continue working with $c=1$. Expanding the Hamiltonian \eqref{Hgvsr1} in powers of $\nicefrac{1}{m}$, we obtain its version up to $\nicefrac{1}{m^3} -$corrections 
\begin{eqnarray} \label{Hgvsr2}
    \mathcal{H}^g_{VSR} &\simeq& (m+\frac{\lambda}{m}) V\gamma^0 - i (1+\frac{\lambda}{m^2})\gamma^0 \gamma^i  (V \partial_i  +  \frac12 \partial_i V   )\nonumber\\
    && - i \frac{\lambda}{m^2} \gamma^0\gamma^j \tilde n^i \tilde n_j (V\partial_i +\frac12 \partial_i V ) + i \frac{\lambda}{2 m^2} \sigma^{ij } \tilde n_j \partial_i V \nonumber\\
    && - \frac{\lambda }{m^3}  \gamma^0 (\tilde n^i \tilde n^j + \eta^{ij} )(V \partial_i \partial_j + \partial_i V \partial_j) \nonumber\\
    && - \frac{\lambda}{m^3 } \gamma^k \tilde n_k (\tilde n^i \tilde n^j + \eta^{ij} ) \partial_i V \partial_j\,.
\end{eqnarray}
We should stress that up to now we have only neglected terms based on their $\nicefrac{1}{m}-$ and $\lambda-$order. In the following, we will also discard them based on their PN$-$order according to the previous discussion in Section \ref{sec:orderofpert}. To proceed, we must identify the even and odd operators from \eqref{Hgvsr2}
\begin{eqnarray} \label{EOvsrcasegrav}
    \mathcal E^g &=& m v \gamma^ 0 + \frac{\lambda}{m} (1+v)\gamma^0  - \frac{\lambda}{m^3} \gamma^0 ((\tilde n^i \partial_i )^2 +\partial_i \partial^i  )  \,, \nonumber \\
    \Theta^g &=& - i (1+\frac{\lambda}{m^2})\gamma^0 \gamma^i \partial_i -i \frac{\lambda}{m^2} \gamma^0\gamma^i \tilde n^j \tilde n_i \partial_j \,. 
\end{eqnarray}
The above expressions are already approximated up to 0PN. Moreover, we have excluded $\nicefrac{1}{m^3}-$terms from $\Theta^g$, because when included in the calculations of the relevant structures in \eqref{commstructure} they produce either:
\begin{itemize}
    \item Even terms at most $\nicefrac{1}{m^4}$. 
    \item Odd terms at most $\nicefrac{1}{m^3}$, but higher-order in other parameters, such as $\lambda $ and $\nicefrac{1}{c}$. 
\end{itemize}
Therefore, repeating this argument until necessary, all contributions produced by those initial odd $\nicefrac{1}{m^3}-$terms become irrelevant. \\
The form of the Hamiltonian after the first transformation can be compactly written as the sum
\begin{eqnarray} \label{HFWafter}
    \mathcal H ' &\simeq& \gamma^0 m c^2 + \mathcal{E}^g +\frac{1}{2m } \gamma^0   [\Theta^g, \mathcal{E}^g]+ \frac{1}{2m } \gamma^0 (\Theta^g)^2 \nonumber\\
    && -\frac{1}{8 m^2 } [\Theta^g, [\Theta^g,\mathcal{E}^g]] +...
\end{eqnarray}
In principle, we should also have considered terms involving time derivatives of the operator $S$, as seen in \eqref{HFWcommut}. However, being
\begin{equation}
    \dot S = - \frac{1}{m } \gamma^0 \dot \Theta^g =  i \frac{\lambda}{m^3} \gamma^i \partial_t(\tilde n^j \tilde n_i )\partial_j 
   \,\,\propto \,\,\nicefrac{1}{c^2} \,,
\end{equation}
those contributions belong to higher PN$-$orders and are then negligible. Calculating each term in \eqref{HFWafter} up to the corresponding accuracy level, we find 
\begin{eqnarray}\label{commstructure}
    \, [\Theta^g, \mathcal{E}^g] &=& -i \frac{2\lambda}{m} \gamma^i \partial_i \,\nonumber\\ 
    \, (\Theta^g)^2 &=& (1+\frac{2\lambda}{m^2})\partial^i\partial_i + \frac{2\lambda}{m^2} (\tilde n^i \partial_i )^2 \,, \\
    \, [\Theta^g, [ \Theta^g, \mathcal{E}^g ]] &=& \frac{4\lambda}{m} \gamma^0 \partial_i \partial^i   \,,\,\, (\Theta^g)^3 = 0 \,, \nonumber\\
    \, 0 &=& (\Theta^g)^4 =  \mathcal{E}^g (\Theta^g)^3   \,. \nonumber
\end{eqnarray}
Replacing these expressions into \eqref{HFWafter}, after a few cancellations, we end up with
\begin{equation}
    \mathcal{H}' = (m+\frac{\lambda}{m}) V \gamma^0 + \gamma^0(1 -\frac{\lambda }{m^2} )\frac{\partial^i \partial_i }{m} -i \frac{\lambda}{m^2} \gamma^0 \gamma^i \partial_i \,.
\end{equation}
Performing a second FW transformation has the only effect of removing the remaining odd terms. Hence, it is straightforward to write down the final 0PN non-relativistic VSR Hamiltonian for UCNs in an accelerated frame
\begin{equation} \label{res0pn}
     H^{g\,\, \text(0PN)}_{VSR} = \gamma^0 \left ( (m+\frac{\lambda}{m})(1+ \vec a \cdot \vec x ) + \frac{\partial_i \partial^i}{2m} (1-\frac{\lambda}{m^2})  \right ) .
\end{equation}
Given that at this order in $\lambda $, the effective mass from Eq.~\eqref{vsrmassf} is $m_f\sim m + \lambda/m$, we can re-express this Hamiltonian to better expose its analogy with \eqref{HgFWlicase},
leading to
\begin{equation} \label{finalaccham}
    H^{g\,\, \text(0PN)}_{VSR} = \gamma^0 \left ( m_f+ m_f \vec a \cdot \vec x + \frac{1}{2m_f} \partial_i \partial^i \right ) .
\end{equation}
Thus, at leading order, there are no VSR modifications in accelerated frames apart from the mass shift previously seen. Even though we could have expected the lack of non-trivial VSR terms at 0PN from dimensional arguments, this outcome is still noteworthy as it reveals at least three fundamental aspects: The absence of differences among particle-antiparticle sectors, the validity of the equivalence between inertial and gravitational mass (now in terms of the effective mass $m_f$), and lastly the fact that in the NR world ($c\to\infty$) the preferred direction does not appear to play a role. Those features were not at all evident a priori from the relativistic EOM \eqref{vsreomgrav}.\\
Nevertheless, this result does not allow us to place constraints on the VSR parameters by using observations from gravitational spectroscopy experiments, such as \emph{q}\textsc{Bounce}. That is because, at 0PN order, $\lambda $ enters the non-relativistic expressions just to correct the particle mass from its Dirac value $m$ to the effective one $m_f$. Then, since the neutron mass is an input parameter obtained from other experiments, we have no way to disentangle its Dirac and VSR components at LO. From that comes the necessity of a higher-order approximation.

\subsection{Next-to-Leading Order -- $ c^{-1}$}

We now want to go further in the NR expansion and compute the NLO corrections at 0.5PN. Taking into account the discussion in Section \ref{NLOsec}, the relevant even and odd operators for this case are
\begin{eqnarray} \label{EOvsrcasegrav2}
    \mathcal E^g &=& m v \gamma^ 0 + \frac{\lambda}{m} (1+v)\gamma^0 +i \frac{\lambda}{2m^2} \sigma^{ij } \tilde n_j \partial_i v \,, \\
    \Theta^g &=& - i (1+v)\gamma^0 \gamma^i \partial_i - \frac{i}{2} \gamma^0 \gamma^i \partial_i v \,, \nonumber
\end{eqnarray}
where we already excluded all $\nicefrac{1}{m^3}-$terms from \eqref{Hgvsr2}, together with the odd $\nicefrac{1}{m^2}-$ones. Note that expressions proportional to $\dot S$ are once again irrelevant for our expansion. Indeed, they contain at most odd $\nicefrac{1}{m^3}-$ or even $\nicefrac{1}{m^4}-$contributions, which are both negligible here. Therefore, the appropriate formula for the transformed Hamiltonian is still \eqref{HFWafter}. Repeating the calculations for the required structures, we get
\begin{eqnarray} \label{0.5pnhp}
    \mathcal{H}' &=& m(1+ \frac{\lambda}{m^2})V \gamma^0 +\gamma^0 \frac{\partial^i\partial_i}{2m}+i\frac{\lambda}{2m^2} \sigma^{ij}\tilde n_j \partial_i v \nonumber\\
    && - i v \gamma^0 \gamma^i \partial_i -\frac{i}{2} \gamma^0 \gamma^i \partial_i v \,. 
\end{eqnarray}
Exactly as in the 0PN case, the effect of performing a second FW transformation is just to eliminate the odd part of \eqref{0.5pnhp}. Thus, we can directly read off the expressions of the new 0.5PN operator 
\begin{eqnarray}
    \delta H^{g\,\, \text(0.5PN)}_{VSR} = -i\frac{\lambda}{2m^2} \sigma^{ij}\tilde n_i \partial_j v \,.
\end{eqnarray}
Naturally, the even part of \eqref{0.5pnhp} is incomplete since, by truncating the expansion at $\nicefrac{1}{m^2}$, we are missing out on some 0PN components. Consequently, the correct way to obtain the final outcome up to 0.5PN order is to merge the result in \eqref{res0pn} with the above corrections, obtaining
\begin{eqnarray}
    H^{g\,\, \text(0.5PN)}_{VSR} &=& H^{g\,\, \text(0PN)}_{VSR} + \delta H^{g\,\, \text(0.5PN)}_{VSR} \\
    &=& (m+ \frac{\lambda}{m})( 1+v)\gamma^0 + \frac{\gamma^0}{2m} (1 - \frac{\lambda}{m^2}) \partial_i \partial^i \nonumber \\
    && -i\frac{\lambda}{2m^2} \sigma^{ij}\tilde n_i \partial_j v  \,, \nonumber
\end{eqnarray}
or in terms of the shifted mass $m_f$ and the explicit content of $v= \vec a \cdot \vec x$
\begin{eqnarray} \label{finalresultH}
    H^{g\,\, \text(0.5PN)}_{VSR} &=& \gamma^0 \left(m_f + m_f \vec a \cdot \vec x  + \frac{1}{2m_f} \partial_i \partial^i \right) \nonumber \\
    &&+i\frac{\lambda}{2m_f^2} \sigma^{ij}\tilde n^i a^j  \,. 
\end{eqnarray}
Contrary to what happened in the 0PN expansion, we now observe the existence of non-trivial VSR corrections, which involve the special direction $\vec {\tilde n}$. Those can then be used to experimentally test the VSR hypothesis.

\section{Comparison with Experiments} \label{expconn}

At this point, we are ready to connect our previous results with \emph{q}\textsc{Bounce}-like experiments and configurations. To do that, we start by recognizing in \eqref{finalresultH} an unperturbed Hamiltonian $H_0$ (equivalent to the one in \eqref{HgFWlicase}) and a perturbation $\Lambda $ originating from VSR
\begin{eqnarray}
    H_0 &=&  m_f + m_f \vec a \cdot \vec x  + \frac{1}{2m_f} \partial_i \partial^i \,,  \\
    \Lambda &=&   \frac{\lambda}{2m^2_f} \vec \sigma \cdot (\vec {\tilde n} \times \vec a)  \,, \label{lambdaexpr}
\end{eqnarray}
where we have specialized to matter particles, taking into consideration that
\begin{equation}
    \sigma^{ij } \tilde n^i a^ j = - i \epsilon^{ijk} \Sigma ^k \tilde n^i a^j \underset{\text{matter}}{\Longrightarrow} - i \epsilon^{ijk} \sigma ^k \tilde n^i a^j \,.  
\end{equation}
We should stress that the single term in $\Lambda$ is analogous to the spin-orbit coupling in the electromagnetic case, where the spin operator is represented by $\vec S = \frac{\hbar }{2} \vec \sigma $.\\
As seen in \eqref{acctime}, the timescale $t_a$ of the $\vec{\tilde n}-$evolution is so large that the vector $\vec {\tilde n}$ can effectively be considered constant over the duration ($\sim 10^{-2}\,s$) of each measurement. Thus, we can employ time-independent and degenerate quantum perturbation theory to calculate the 0.5PN corrections to the unperturbed energy levels defined by $H_0$. To do that, we first calculate the matrix elements $\Lambda^N _{\alpha\beta}$ of the operator $\Lambda$ in each degenerate eigenspace
\begin{equation}
    \Lambda^n_{s_1 s_2} = \bra{N , s_1}  \Lambda \ket{N, s_2} \,,
\end{equation}
where $N$ is the quantum number labeling the unperturbed spectrum and $s_1, s_2 = \,\uparrow , \downarrow $ represents the spin projection over the quantization direction, which in this case is the one of the acceleration, i.e., $\vec{\hat z}$. Recalling $\vec a = (0,0,a)$, we use the relation
\begin{equation}
    \vec \sigma \cdot (\vec {\tilde n} \times \vec a) = a \left( \begin{array}{cc} \;  0 & \tilde n^2 + i\tilde n^1 \\ \tilde n^2 - i \tilde n^1 & \; 0 \end{array}\right) ,
\end{equation}
and the information on the unperturbed orthonormal energy levels $\ket{N, s}$ of the quantum bouncing ball \cite{PhysRevD.109.064085}, to find
\begin{equation}
    \Lambda^N_{s_1 s_2} = \frac{a\lambda}{2 m^2_f} \left( \begin{array}{cc} \;  0 & \tilde n^2 + i\tilde n^1 \\ \tilde n^2 - i \tilde n^1 & \; 0 \end{array}\right) .
\end{equation}
Finally, setting to zero the determinant of $ \Lambda - \varepsilon \, \mathbf{1}_{2\times 2 }$, we obtain the value of the energy corrections 
\begin{equation} \label{varnepsilon}
   { \varepsilon_N} = \pm \frac{a\lambda}{2m^2_f} \sin\theta \,,
\end{equation}
which, remarkably, is independent of $N$, meaning that all energy levels are split in the same way. \\
Let us now present a rough estimate for the simplest case $\theta=\nicefrac{\pi}{2}$, assuming the neutron mass $m_f \simeq 939.565\,\nicefrac{ MeV}{c^2} $ \cite{Mohr:2024kco} and the acceleration value $a= 9.8049 \,m/s^2$ measured next to the \emph{q}\textsc{Bounce} experiment utilizing a falling corner cube \cite{Micko:2023oar}. If we imagine an experiment with the same sensitivity reached by \emph{q}\textsc{Bounce}, which is at the moment around $\sim 10^{-16}\,eV$, deviations arising from VSR could be detected for values of $\sqrt{|\lambda|} \gtrsim 1 \, { TeV}/{c^2} $. Nevertheless, the latter is evidently a range far outside our assumption $|\lambda| \ll  m^2\simeq m^2_f $. Hence, additional clever ideas should be implemented in the future to increase \emph{q}\textsc{Bounce} sensitivity to this particular effect.\\
Interestingly enough, it is possible to classify the correction \eqref{lambdaexpr} according to the notation implemented by Kosteleck\'y and collaborators in their work \cite{Kostelecky:2021tdf}, by simply identifying
\begin{equation}\label{kNRsg}
    (k^{NR}_{\sigma g})^{ij } = \frac{\lambda}{2 m^2_f} \epsilon^{ikj} \tilde n^k 
    \,.
\end{equation}
However, the bounds found in Ref.~\cite{Kostelecky:2021tdf} for the trace of such LV tensor are not helpful here, since the trace of \eqref{kNRsg} vanishes exactly.

\section{Discussion and Conclusions} \label{chap5}

In this work, we have taken an initial step toward analyzing the gravitational phenomenology of VSR fermionic systems by computing the leading and next-to-leading order corrections to the non-relativistic Hamiltonian in an accelerated frame. That is equivalent to the scenario of a uniform and homogeneous gravitational field, which can be considered as the first approximation to real gravitational backgrounds. \\
The absence of non-trivial modifications observed in the 0PN Hamiltonian further demonstrate the elusive nature of peculiar LV realizations like VSR. The new corrections found at 0.5PN depend instead on the preferred direction introduced in the VSR algebra. Therefore, they can be used to probe novel Lorentz-violating signatures and set new constraints on the VSR parameter for the neutron sector. However, current gravitational spectroscopy experiments, such as \emph{q}\textsc{Bounce}, only measure spin-conserving transitions, which seem not sensible to the VSR perturbations in the 0.5PN spectrum. Thus, new experimental configurations are needed to enable the investigation of these effects.\\
Another interesting aspect that emerged from our study is the peculiar time-dependent behavior of the preferred spatial direction $\vec{\tilde n}$ in accelerated frames. In fact, while the presence of acceleration is what allows for non-trivial VSR terms, we have also shown that the greater the acceleration, the more rapidly $\vec{\tilde n}$ tends to anti-align with $\vec a$, making the new VSR correction in \eqref{varnepsilon} vanish and weakening the relevance of initial conditions. Furthermore, this evolution should also depend on the underlying nature that we assume for VSR. Indeed:
\begin{itemize}
    \item If $n^\mu$ emerges from some new internal degree of freedom, its evolution might occur only during measurements, or more generally, during the neutrons' lifetimes, as long as they remain in free fall.
    \item In contrast, if it arises from global spacetime features, as usually considered, its evolution would start together with the appearance of acceleration, implying that in most cases $\vec{\tilde n}$ may have had enough time to anti-align with $\vec a$.
\end{itemize}
One factor that we have neglected in this analysis is the presence of Earth's rotation. The latter could alter the evolution of $\vec{\tilde n} $ because the interplay between the ongoing rotational motion and the tendency to align with $\vec a$ may give rise to a precession pattern. Moreover, when modeling laboratories on Earth's surface, rotational effects are certainly relevant, since they can naturally lead to extra 0.5PN corrections. For these reasons, we wish to overcome this limitation in the future.

\begin{acknowledgments}

The authors would like to thank the referee for the insightful response, which allowed for an improved revision of this work. A.S. acknowledges financial support from the ANID CONICYT-PFCHA/DoctoradoNacional/2020-21201387. E.M. acknowledges financial support from Fondecyt Grant No 1230440.\\

\end{acknowledgments}

\appendix

\section{Geometric Quantities for Accelerated Observers} \label{app:rindlerquantities}

Here, we include the expressions of several geometric quantities related to the spacetime metric in \eqref{rindlerelement}. Recalling that $V\equiv 1+\frac{\vec a\cdot \vec x }{c^2}$ with $\vec a $ constant and homogeneous, we first list the non-zero components of the Christoffel symbols
\begin{eqnarray} \label{christsymbols}
    \{^{\;\, 0}_{ 0 \,\, i}\} &=& \frac{1}{V}\partial_i V \,,\\
    \{^{\;\, i}_{ 0 \,\, 0}\} &=& V \partial_i V\,. \nonumber
\end{eqnarray}
As already mentioned in the main text, the components of the Riemann tensor are, instead, all zero. Then we move on to the calculation of the components of the spinor connection \eqref{spinorconn}. Starting from the temporal one, we have
\begin{eqnarray}
    \Gamma_0 &=& \frac{1}{4} \sigma^{ab} g_{\alpha \beta} E_a^{\;\;\alpha} (\partial_t E_{ b}^{\;\;\beta} +\{^{\;\, \beta}_{ 0\, \rho}\} \, E_{ b}^{\;\;\rho}) \nonumber\\
    &=& \frac{1}{4} \sigma^{ab} g_{\alpha \beta} \{^{\;\, \beta}_{ 0\, \rho}\} E_a^{\;\;\alpha}  \, E_{ b}^{\;\;\rho} \,.
\end{eqnarray}
Using the expressions \eqref{christsymbols}, we obtain
\begin{eqnarray}
    \Gamma_0 &=& \frac{1}{4} \sigma^{0i} g_{00} \{^{\;\, 0}_{ 0\, \,i}\} E_0^{\;\;0}  \, E_{ i}^{\;\;i}+ \frac{1}{4} \sigma^{i0} g_{ii} \{^{\;\, i}_{ 0\,\, 0}\} E_i^{\;\;i}  \, E_{ 0}^{\;\;0} \nn \\
    &=& \frac{1}{4V^2} \sigma^{0i} V^2 \partial_i V - \frac{1}{4V} \sigma^{i0} V \partial_i V \\ 
    &=& \frac{1}{2} \sigma ^{0i} \partial_i V \,, \nonumber
\end{eqnarray}
where we also applied the formulas \eqref{rindlertetrads} for the tetrads. Finally, using the properties of flat gamma matrices, we end up with
\begin{equation} \label{gammazero}
    \Gamma_0 = \frac12 \sigma^{0i} \partial_i V = \frac{1}{2} \gamma^0 \gamma^i \partial_i V\,.
\end{equation}
Repeating the same steps for the spatial components of $\Gamma_\mu$ we arrive, instead, at
\begin{equation}
    \Gamma_i = 0\,,
\end{equation}
meaning that only the time component of the spinor covariant derivative $\mathcal D_0 \psi$ gets a modification from the non-inertial geometry.

\section{Evolution of $\tilde n^ i$} \label{app:evoofn}

Consider the differential equation \eqref{timeevofn} that determines the time evolution of the unit vector $\vec {\tilde{n}}$
\bea
\partial_t \tilde{n}^i = \tilde{n}^i (\vec{\tilde{n}}\cdot\vec{a}) - a^i  \,,
\label{eq_diffn}
\eea
where we remember that $ a^i  = \partial_i V$ is the laboratory constant acceleration vector, which is equal and opposite to the local gravitational acceleration $\vec g$. To mimic conditions analogous to those on Earth's surface, we choose the reference frame such that the acceleration points along the $\hat{z}$-axis, i.e.
\be
\vec{a} = (0,0,a) \,\,,\,\,\, \text{with } a >0 \, .
\ee
Moreover, since $\vec{\tilde{n}}$ is a unit vector, it is convenient to parameterize it in spherical coordinates
\bea
\vec {\tilde{n}} = \left( \sin\theta \cos\varphi,\sin\theta\sin\varphi,\cos\theta \right) .
\eea
Hence, we have
\be
\vec{\tilde{n}}\cdot\vec{a} = a\cos\theta \, .
\ee
Employing these choices of frame and parameterization, the differential equation system~\eqref{eq_diffn} reduces to
\bea\label{eq_diffang}
\partial_t\left(\sin\theta\cos\varphi\right) &=& a \sin\theta \cos\theta \cos\varphi \,, \nonumber \\
\partial_t\left( \sin\theta\sin\varphi \right) &=& a \sin\theta \cos\theta \sin\varphi \,, \\
\partial_t\left(\cos\theta\right) &=& a \cos^2\theta -a \nonumber \,.
\eea
We start by explicitly solving the third differential equation in \eqref{eq_diffang}. Defining the variable $y(t) \equiv \cos\theta(t)$ and assuming the initial condition $y(0) = \cos\theta_0$, we get
\bea
y(t) = \cos\theta(t) = - \frac{1 - \beta e^{-2 a t}}{1 + \beta e^{-2 a t}} \,,
\label{eq_costheta}
\eea
where we introduced the parameter
\be
\beta \equiv \frac{1 + y(0)}{1 - y(0)} = \frac{1 + \cos\theta_0}{1 - \cos\theta_0} \,.
\label{eq_beta}
\ee
Let us now turn to the solution of the first differential equation. Using the auxiliary variable $u(t) \equiv \sin\theta(t)\cos\varphi(t)$, we obtain
\bea
\frac{du}{dt} = a u(t) y(t) \,,
\label{eq_diffequ}
\eea
subject to initial condition $u(0) = \sin\theta_0\cos\varphi_0 =u_0$. Eq.~\eqref{eq_diffequ} can be solved by separation of variables, thanks to the knowledge of the explicit solution for $y(t) $ in Eq.~\eqref{eq_costheta}. The final result is
\begin{equation} \label{formulaut}
    u(t) = u_0 (1+\beta ) \frac{e^{-a t }}{1 + \beta e ^{-2 a t}} \,.
\end{equation}
To find the explicit behavior on the equatorial plane we observe that
\begin{eqnarray}
    \sin\theta(t) &=& \sqrt{1-y(t)^2} = \frac{2 \sqrt{\beta} e^{-a t}}{1+\beta e^{-2a t}} \,, \nonumber \\
    \sin\theta_0 &=& \frac{2 \sqrt{\beta} }{1+\beta } \,.
\end{eqnarray}
Then, replacing it back in \eqref{formulaut}, we derive
\begin{eqnarray}
    \cos\phi(t) &=&  \cos\phi_0 (1+\beta ) \frac{\sin\theta_0}{\sin\theta(t)} \frac{e^{-a t }}{1 + \beta e ^{-2 a t}} \nonumber \\
    &=&  \cos\phi_0 \,,
\end{eqnarray}
implying that the $\phi -$angle remains constant over time. Assuming $a = 9.8049 \,m/s^2$ as in the main text and defining the acceleration time scale $t_a$ as 
\begin{eqnarray}\label{acctime}
    t_a = \frac{c}{2a} \sim 10^ 7 s \sim 115 \,\,days \,,
\end{eqnarray}
the time-dependency of the azimuthal angle can instead be described by
\begin{equation}
    \theta(t) = \cos^{-1} \left (- \frac{ \frac{1}{\beta} -e^{ -t /t_a} }{\frac{1}{\beta} +e^{-t/t_a} } \right ) \,,
\end{equation}
from which it becomes clear that $\theta =\pi$ is a fixed point. Thus, during the evolution of the system, the space direction labelled by $\vec{\tilde n}$ tends to anti-align with the local acceleration vector $\vec a$. The only exception to that is for $\theta_0 =0$ which looks like a point of unstable equilibrium, in the sense that it is stable over time evolution, but any small perturbation would lead to a flip of the $\vec{\tilde n} -$orientation.

\subsection{Kottler-Møller Coordinates and Constant Vectors }

One of the possible charting of a flat spacetime is given by Kottler-Møller coordinates, which naturally adapt to accelerating observers. Starting from a Minkowskian coordinate system $X^\mu = \{ T,X,Y,Z\}$ and taking the acceleration $\vec a// \vec{\hat z}$, we can define the new curvilinear one $x^\mu = \{ t,x,y,z\}$ through the following definitions 
\begin{eqnarray} \label{kotmol}
t &=& \frac{1}{a} \arctanh{\left( \frac{T}{Z+\frac{1}{a}}\right ) } \,, \nonumber\\
x &=& X \,, \\
y &=& Y \,, \nonumber \\
z &=& \sqrt{ \left ( Z + \frac{1}{a}\right)^2 - T^2 } - \frac{1}{a} \,,\nonumber
\end{eqnarray}
which leads to a spacetime element $ds^2$ analogous to the one seen in \eqref{rindlerelement}
\begin{equation}
    ds^2 = dT^2 - d\vec X ^2 = (1+az)^2 dt^2 -d\vec x^2 \,.
\end{equation}
The Jacobian transformation matrix which connects the two coordinate charts is given by
\begin{equation} \label{kotlambda}
    \Lambda_{\nu}^{\;\mu} (x) = \frac{\partial x^\mu}{\partial X^\nu } =
    \begin{pmatrix}
        \frac{\cosh a t  }{1+az} & 0 & 0 & - \frac{\sinh at }{1+az} \\ 
    0 & 1 & 0 & 0 \\
    0 & 0 & 1 & 0 \\
    -\sinh at & 0 & 0 & \cosh at
    \end{pmatrix} .
\end{equation}
At this point, the peculiar time dependence of $\tilde n^i$ can be directly explained using the Kottler-Møller transformation \eqref{kotmol}: Let us start from a lightlike vector $v^\mu = (v^0 , \vec v)$, constant in the old inertial coordinates $\partial_{X^\mu} v^\nu = 0$. Transforming its components to the new coordinates using \eqref{kotlambda}, we derive
\begin{eqnarray}
    v^{\mu' } &=& \Lambda_{\mu}^{\;\mu'} v^\mu \,, \\
    v^t &=& \frac{v^T\cosh at  - v^Z \sinh at }{1+ az} \,,\nonumber \\
    v^{x^i} &=& ( v^X, v^Y, v^Z \cosh at - v^T\sinh at) \,. \nonumber
\end{eqnarray}
Now, defining for $ v^\mu$ a quantity analogous to $\vec{\tilde n}$ in the main text, we obtain
\begin{equation}
    \tilde v^{x^i} \equiv \frac{v^{x^i}}{v^t (1+az)} \,,
\end{equation}
the components of which are given by
\begin{eqnarray} \label{tildevi1}
    \tilde v^{x^i} &=& \frac{1}{ v^T \cosh at - v^Z \sinh at} \times \\
    && \times \,(v^X, v^Y, v^Z \cosh at  - v^T \sinh at) \nonumber\,.
\end{eqnarray}\\
Finally, considering \eqref{tildevi1} for very long times $t$, we get
\begin{equation}
  \lim_{t\to\infty}  \tilde v^{x^i} = ( 0,0,-1 ) \,,
\end{equation}
meaning that ${ \tilde v^{x^i}}$ tends to anti-align with the acceleration, exactly as it happened for $\tilde n^i$. Once again, the only exception to that is for $\vec { v} // \vec a$, since it implies $ v^Z = v^T$ and then $\tilde v^{x^i}$ is just constantly equal to
\begin{eqnarray}
    \tilde v^{x^i} = (0,0,1) \,. 
\end{eqnarray}
Thus, this is another way to think about the origin of the time-dependent behavior of $\tilde n^ i$.

\section{Hermiticity Concerns and Spinorial Metrics} \label{app:spinmetric}

In this appendix, we include more details on the strategy followed to determine the spinorial metric $ G$ in VSR and its respective decomposition. 

\subsection{Adjoint of the VSR Hamiltonian}

Initially, we must establish that \eqref{Hgvsr1.5} is not Hermitian by calculating its adjoint. The latter is defined by the relation
\begin{equation}
    (\mathcal H^g_{VSR}  \psi, \psi )= (\psi , (\mathcal H^g_{VSR})^\dagger \psi) \,.
\end{equation}
In order to move the non-local operators in the Hamiltonian from one spinor to the other, we first observe that  
\begin{equation}
   \tilde{\mathcal R}_\partial V \psi = V \tilde{\mathcal R}_\partial \psi \,- [\tilde{ r}_\partial, V ] \tilde{\mathcal R}^2_\partial \psi \,.
\end{equation}
Moreover, exploiting the integral representation for $\tilde{\mathcal R}_\partial$ and integration by parts, we note that
\begin{equation}
    \int d^3x \, \psi \, \tilde{\mathcal R}_\partial \, \varphi = \int d^3x \, (\tilde{\mathcal R}_\partial \psi ) \, \varphi \,.
\end{equation}
Defining the operator
\begin{equation}
    \mathcal K \equiv (1 -\gamma^0 \gamma^i \tilde n_i) \tilde{\mathcal R}_\partial\,,
\end{equation}
and dividing the Hamiltonian \eqref{Hgvsr1.5} into its parts at zero and first order in $\lambda$, we respectively obtain
\begin{eqnarray}
   \mathcal  H_{(0)} &=& mV\gamma^0-iV \gamma^0 \gamma^i \partial_i - \frac{i}{2} \gamma^0 \gamma^ i \partial_i V \,, \\
    \mathcal H_{(1)} &=& V \mathcal K \,( m \gamma^0 - i \gamma^0 \gamma^i \partial_i + i \tilde n^ i \partial_i )  \nonumber \,.
\end{eqnarray}
It is straightforward to see that, while $ \mathcal H_{(0)}$ is Hermitian, $ \mathcal H_{(1)} $ is not. In fact, we have
\begin{equation}
    \mathcal H_{(1)}^\dagger = \mathcal K \,V ( m \gamma^0 - i \gamma^0 \gamma^i \partial_i + i \tilde n^ i \partial_i )  + i \mathcal K (\tilde n^ i -\gamma^0 \gamma^i  ) \partial_i V \,.
\end{equation}
This implies, as anticipated, that
\begin{equation}
    (\mathcal H^g_{VSR})^\dagger = \mathcal H_{(0)} + \mathcal H _{(1)}^\dagger \neq \mathcal H^g_{VSR} \,.
\end{equation}

\subsection{Spinorial Metric $G$ in VSR}

Our starting assumption for the calculation of $G$ is 
\begin{equation} \label{gplusdg}
    G = 1 + \frac{\lambda}{m^2} G_{(1)} \,,
\end{equation}
which is based on the following two observations:
\begin{itemize}
    \item In the limit $\lambda\to0$, we must recover a trivial spinorial metric $G=1$.
    \item Since the Hamiltonian \eqref{Hgvsr1.5} contains up to linear terms in $\lambda$, we expect the same for $G$.
\end{itemize}
Then, matching terms of equal order in $\lambda$, the defining relation \eqref{definingrelG} for the spinorial metric gets split into two separate ones
\begin{equation}
\begin{cases}
    \partial_t G_{(1)} + i\, [\mathcal H_{(0)}, G_{(1)}] + i \left ( \mathcal H^\dagger_{(1)} - \mathcal H_{(1)} \right )  =0\,,\\
    \\
    [\mathcal H_{(1)}, G_{(1)}] + \left (\mathcal H^\dagger_{(1)}-\mathcal H_{(1)} \right ) G_{(1)}  =0 \,.
\end{cases}
\end{equation}
At this point, it is straightforward to check that taking $G_{(1)}= \mathcal K$ both of the conditions above are satisfied, so that \eqref{Gexpr} is really the correct solution of \eqref{definingrelG}.

\subsection{Metric Decomposition}

In order to achieve the decomposition of $G=\Delta^\dagger \Delta$, we try searching for a $\Delta$ structured as follows
\begin{equation} \label{deltab1b2}
    \Delta = b_1 + b_2 \gamma^0 \gamma^i \tilde n_i \,, 
\end{equation}
which we also assume to be Hermitian. Therefore, from $G=\Delta^2$, we obtain the relations
\begin{equation}
    \begin{cases}
        (b_1)^2 + (b_2)^2 =1 + \frac{\lambda}{m^2} \tilde{\mathcal R}_\partial \,, \\
         2 b_1 b_2= -\frac{\lambda}{m^2} \tilde{\mathcal R}_\partial \, \to\, b_2 = - \frac{\lambda}{2 \,b_1 m^2} \tilde{\mathcal R}_\partial  \,.
    \end{cases}
\end{equation}
Since we want $\Delta $ to satisfy $\lim _{\lambda\to 0} \Delta =1$, meaning $\lim _{\lambda\to 0} b_1 =1$ and $\lim _{\lambda\to 0} b_2 =0$, a possible solution when solving for $b_1$ is  
\begin{equation}
    b_1 = \sqrt{\frac{1+ \frac{\lambda}{m^2} \tilde{\mathcal R}_\partial +\sqrt{1 +\frac{2\lambda}{m^2} \tilde{\mathcal R}_\partial}}{2}} \,,
\end{equation}
and consequently
\begin{equation}
    b_2 =-\frac{\frac{\lambda}{m^2} \tilde{\mathcal R}_\partial}{2^\frac12 \sqrt{1+ \frac{\lambda}{m^2} \tilde{\mathcal R}_\partial +\sqrt{1 +\frac{2\lambda}{m^2} \tilde{\mathcal R}_\partial}}} \,.
\end{equation}
Plugging these formulas into \eqref{deltab1b2}, we get to the expression \eqref{deltaexprexact} for $\Delta$ in the main text.

\subsection{Positive-Definiteness of $G$} \label{app:posdef}

The positivity of the spinor norm induced by the generalized inner product requires to have
\begin{equation} \label{posdef1}
    (\psi ,\psi)_G > 0 \,\,\,\,\, ,\,\, \forall\psi \neq0 \,.
\end{equation}
To prove the latter statement, we work in momentum space to avoid unnecessary complications. In fact, by means of Fourier transformations, since the metric $G$ does not contain $\vec x-$dependent coefficients, we can rewrite the scalar product as
\begin{equation}
    (\psi ,\psi)_G = \int \frac{d^3p}{(2\pi)^3} \, \hat \psi^\dagger(\vec p) \hat G (\vec p) \hat \psi(\vec p) \,,
\end{equation}\\
where we use the “hats” to denote the Fourier transforms of the spinors and the metric. Thus, the condition \eqref{posdef1} reduces to
\begin{equation}
    \hat \psi^\dagger(\vec p) \hat G (\vec p) \hat \psi(\vec p) > 0 \,\,\,\,\, , \,\, \forall \vec p \,,\,\, \forall \hat \psi(\vec p)\neq0 \,,
\end{equation}
which can be ensured by determining the positivty of the eigenvalues of $\hat G$. Defining the momentum $\vec p_\perp$ perpendicular to $\vec{\tilde n}$ as
\begin{equation}
    \vec p _\perp \equiv \vec p - (\vec{\tilde n }\cdot \vec p) \; \vec{\tilde n}\,,
\end{equation}\\
we can write $\hat G (\vec p) $ as
\begin{equation}
   \hat G = 1+  \frac{\lambda}{m^2 + \vec p ^2_\perp} (1 - \gamma^ 0 \gamma^i \tilde n_i) \,.
\end{equation}
Therefore, thanks to the algebra of Dirac matrices, we can easily calculate the determinant of $\hat G - \alpha \, Id_4$ and find the eigenvalues of $\hat G$ by solving 
\begin{eqnarray}
    0 &=& \left ( (1 + \frac{\lambda}{m^2+\vec p ^2_\perp} -\alpha )^2 - (\frac{\lambda |\vec{\tilde n}| }{m^2 +\vec p_\perp^2} )^2 \right )^2 \nonumber \\
    &=& \left (1 + \frac{2\lambda}{m^2+\vec p ^2_\perp}-2 \alpha \, (1+\frac{\lambda}{m^2+\vec p ^2_\perp}) + \alpha^2 \right )^2 \nonumber . \\
\end{eqnarray}
The result is straightforward
\begin{equation}
    \alpha = \begin{cases}
        1 \,, \\
        1 + \frac{2\lambda}{m^2 + \vec p^2_\perp} \,.
    \end{cases}
\end{equation}
Both solutions are clearly always positive for $\lambda>-m^2/2$, demonstrating the positive-definiteness of the spinorial metric $G$ and its respective norm, as expected.



\bibliography{./biblio/apssamp}

\providecommand{\noopsort}[1]{}\providecommand{\singleletter}[1]{#1}%
\begin{thebibliography}{64}%
\makeatletter
\providecommand \@ifxundefined [1]{%
 \@ifx{#1\undefined}
}%
\providecommand \@ifnum [1]{%
 \ifnum #1\expandafter \@firstoftwo
 \else \expandafter \@secondoftwo
 \fi
}%
\providecommand \@ifx [1]{%
 \ifx #1\expandafter \@firstoftwo
 \else \expandafter \@secondoftwo
 \fi
}%
\providecommand \natexlab [1]{#1}%
\providecommand \enquote  [1]{``#1''}%
\providecommand \bibnamefont  [1]{#1}%
\providecommand \bibfnamefont [1]{#1}%
\providecommand \citenamefont [1]{#1}%
\providecommand \href@noop [0]{\@secondoftwo}%
\providecommand \href [0]{\begingroup \@sanitize@url \@href}%
\providecommand \@href[1]{\@@startlink{#1}\@@href}%
\providecommand \@@href[1]{\endgroup#1\@@endlink}%
\providecommand \@sanitize@url [0]{\catcode `\\12\catcode `\$12\catcode `\&12\catcode `\#12\catcode `\^12\catcode `\_12\catcode `\%12\relax}%
\providecommand \@@startlink[1]{}%
\providecommand \@@endlink[0]{}%
\providecommand \url  [0]{\begingroup\@sanitize@url \@url }%
\providecommand \@url [1]{\endgroup\@href {#1}{\urlprefix }}%
\providecommand \urlprefix  [0]{URL }%
\providecommand \Eprint [0]{\href }%
\providecommand \doibase [0]{https://doi.org/}%
\providecommand \selectlanguage [0]{\@gobble}%
\providecommand \bibinfo  [0]{\@secondoftwo}%
\providecommand \bibfield  [0]{\@secondoftwo}%
\providecommand \translation [1]{[#1]}%
\providecommand \BibitemOpen [0]{}%
\providecommand \bibitemStop [0]{}%
\providecommand \bibitemNoStop [0]{.\EOS\space}%
\providecommand \EOS [0]{\spacefactor3000\relax}%
\providecommand \BibitemShut  [1]{\csname bibitem#1\endcsname}%
\let\auto@bib@innerbib\@empty
\bibitem [{\citenamefont {Collins}\ \emph {et~al.}(2004)\citenamefont {Collins}, \citenamefont {Perez}, \citenamefont {Sudarsky}, \citenamefont {Urrutia},\ and\ \citenamefont {Vucetich}}]{collins2004lorentz}%
  \BibitemOpen
  \bibfield  {author} {\bibinfo {author} {\bibfnamefont {J.}~\bibnamefont {Collins}}, \bibinfo {author} {\bibfnamefont {A.}~\bibnamefont {Perez}}, \bibinfo {author} {\bibfnamefont {D.}~\bibnamefont {Sudarsky}}, \bibinfo {author} {\bibfnamefont {L.}~\bibnamefont {Urrutia}},\ and\ \bibinfo {author} {\bibfnamefont {H.}~\bibnamefont {Vucetich}},\ }\bibfield  {title} {\bibinfo {title} {{Lorentz invariance and quantum gravity: an additional fine-tuning problem?}},\ }\href {https://doi.org/10.1103/PhysRevLett.93.191301} {\bibfield  {journal} {\bibinfo  {journal} {Phys. Rev. Lett.}\ }\textbf {\bibinfo {volume} {93}},\ \bibinfo {pages} {191301} (\bibinfo {year} {2004})},\ \Eprint {https://arxiv.org/abs/gr-qc/0403053} {arXiv:gr-qc/0403053} \BibitemShut {NoStop}%
\bibitem [{\citenamefont {Kosteleck\'y}\ and\ \citenamefont {Samuel}(1989)}]{PhysRevD.39.683}%
  \BibitemOpen
  \bibfield  {author} {\bibinfo {author} {\bibfnamefont {V.~A.}\ \bibnamefont {Kosteleck\'y}}\ and\ \bibinfo {author} {\bibfnamefont {S.}~\bibnamefont {Samuel}},\ }\bibfield  {title} {\bibinfo {title} {Spontaneous breaking of lorentz symmetry in string theory},\ }\href {https://doi.org/10.1103/PhysRevD.39.683} {\bibfield  {journal} {\bibinfo  {journal} {Phys. Rev. D}\ }\textbf {\bibinfo {volume} {39}},\ \bibinfo {pages} {683} (\bibinfo {year} {1989})}\BibitemShut {NoStop}%
\bibitem [{\citenamefont {Kosteleck\'y}(2004)}]{PhysRevD.69.105009}%
  \BibitemOpen
  \bibfield  {author} {\bibinfo {author} {\bibfnamefont {V.~A.}\ \bibnamefont {Kosteleck\'y}},\ }\bibfield  {title} {\bibinfo {title} {Gravity, lorentz violation, and the standard model},\ }\href {https://doi.org/10.1103/PhysRevD.69.105009} {\bibfield  {journal} {\bibinfo  {journal} {Phys. Rev. D}\ }\textbf {\bibinfo {volume} {69}},\ \bibinfo {pages} {105009} (\bibinfo {year} {2004})}\BibitemShut {NoStop}%
\bibitem [{\citenamefont {Mavromatos}(2007)}]{mavromatos2007lorentz}%
  \BibitemOpen
  \bibfield  {author} {\bibinfo {author} {\bibfnamefont {N.~E.}\ \bibnamefont {Mavromatos}},\ }\bibfield  {title} {\bibinfo {title} {{Lorentz Invariance Violation from String Theory}},\ }\href {https://doi.org/10.22323/1.043.0027} {\bibfield  {journal} {\bibinfo  {journal} {PoS}\ }\textbf {\bibinfo {volume} {QG-PH}},\ \bibinfo {pages} {027} (\bibinfo {year} {2007})},\ \Eprint {https://arxiv.org/abs/0708.2250} {arXiv:0708.2250 [hep-th]} \BibitemShut {NoStop}%
\bibitem [{\citenamefont {Carroll}\ \emph {et~al.}(2001)\citenamefont {Carroll}, \citenamefont {Harvey}, \citenamefont {Kostelecky}, \citenamefont {Lane},\ and\ \citenamefont {Okamoto}}]{carroll2001noncommutative}%
  \BibitemOpen
  \bibfield  {author} {\bibinfo {author} {\bibfnamefont {S.~M.}\ \bibnamefont {Carroll}}, \bibinfo {author} {\bibfnamefont {J.~A.}\ \bibnamefont {Harvey}}, \bibinfo {author} {\bibfnamefont {V.~A.}\ \bibnamefont {Kostelecky}}, \bibinfo {author} {\bibfnamefont {C.~D.}\ \bibnamefont {Lane}},\ and\ \bibinfo {author} {\bibfnamefont {T.}~\bibnamefont {Okamoto}},\ }\bibfield  {title} {\bibinfo {title} {{Noncommutative field theory and Lorentz violation}},\ }\href {https://doi.org/10.1103/PhysRevLett.87.141601} {\bibfield  {journal} {\bibinfo  {journal} {Phys. Rev. Lett.}\ }\textbf {\bibinfo {volume} {87}},\ \bibinfo {pages} {141601} (\bibinfo {year} {2001})},\ \Eprint {https://arxiv.org/abs/hep-th/0105082} {arXiv:hep-th/0105082} \BibitemShut {NoStop}%
\bibitem [{\citenamefont {Cohen}\ and\ \citenamefont {Glashow}(2006{\natexlab{a}})}]{vsr1}%
  \BibitemOpen
  \bibfield  {author} {\bibinfo {author} {\bibfnamefont {A.~G.}\ \bibnamefont {Cohen}}\ and\ \bibinfo {author} {\bibfnamefont {S.~L.}\ \bibnamefont {Glashow}},\ }\bibfield  {title} {\bibinfo {title} {{Very special relativity}},\ }\href {https://doi.org/10.1103/PhysRevLett.97.021601} {\bibfield  {journal} {\bibinfo  {journal} {Phys. Rev. Lett.}\ }\textbf {\bibinfo {volume} {97}},\ \bibinfo {pages} {021601} (\bibinfo {year} {2006}{\natexlab{a}})},\ \Eprint {https://arxiv.org/abs/hep-ph/0601236} {arXiv:hep-ph/0601236} \BibitemShut {NoStop}%
\bibitem [{\citenamefont {Cohen}\ and\ \citenamefont {Glashow}(2006{\natexlab{b}})}]{vsr2}%
  \BibitemOpen
  \bibfield  {author} {\bibinfo {author} {\bibfnamefont {A.~G.}\ \bibnamefont {Cohen}}\ and\ \bibinfo {author} {\bibfnamefont {S.~L.}\ \bibnamefont {Glashow}},\ }\bibfield  {title} {\bibinfo {title} {{A Lorentz-Violating Origin of Neutrino Mass?}},\ }\href@noop {} {\  (\bibinfo {year} {2006}{\natexlab{b}})},\ \Eprint {https://arxiv.org/abs/hep-ph/0605036} {arXiv:hep-ph/0605036} \BibitemShut {NoStop}%
\bibitem [{\citenamefont {Alfaro}\ and\ \citenamefont {Soto}(2019)}]{PhysRevD.100.055029}%
  \BibitemOpen
  \bibfield  {author} {\bibinfo {author} {\bibfnamefont {J.}~\bibnamefont {Alfaro}}\ and\ \bibinfo {author} {\bibfnamefont {A.}~\bibnamefont {Soto}},\ }\bibfield  {title} {\bibinfo {title} {Photon mass in very special relativity},\ }\href {https://doi.org/10.1103/PhysRevD.100.055029} {\bibfield  {journal} {\bibinfo  {journal} {Phys. Rev. D}\ }\textbf {\bibinfo {volume} {100}},\ \bibinfo {pages} {055029} (\bibinfo {year} {2019})}\BibitemShut {NoStop}%
\bibitem [{\citenamefont {Alfaro}\ and\ \citenamefont {Santoni}(2022)}]{grav3}%
  \BibitemOpen
  \bibfield  {author} {\bibinfo {author} {\bibfnamefont {J.}~\bibnamefont {Alfaro}}\ and\ \bibinfo {author} {\bibfnamefont {A.}~\bibnamefont {Santoni}},\ }\bibfield  {title} {\bibinfo {title} {{Very special linear gravity: A gauge-invariant graviton mass}},\ }\href {https://doi.org/10.1016/j.physletb.2022.137080} {\bibfield  {journal} {\bibinfo  {journal} {Phys. Lett. B}\ }\textbf {\bibinfo {volume} {829}},\ \bibinfo {pages} {137080} (\bibinfo {year} {2022})},\ \Eprint {https://arxiv.org/abs/2204.05485} {arXiv:2204.05485 [gr-qc]} \BibitemShut {NoStop}%
\bibitem [{\citenamefont {Santoni}\ \emph {et~al.}(2023)\citenamefont {Santoni}, \citenamefont {Alfaro},\ and\ \citenamefont {Soto}}]{Santoni:2023uko}%
  \BibitemOpen
  \bibfield  {author} {\bibinfo {author} {\bibfnamefont {A.}~\bibnamefont {Santoni}}, \bibinfo {author} {\bibfnamefont {J.}~\bibnamefont {Alfaro}},\ and\ \bibinfo {author} {\bibfnamefont {A.}~\bibnamefont {Soto}},\ }\bibfield  {title} {\bibinfo {title} {{Graviton mass bounds in very special relativity from binary pulsar\textquoteright{}s gravitational waves}},\ }\href {https://doi.org/10.1103/PhysRevD.108.044072} {\bibfield  {journal} {\bibinfo  {journal} {Phys. Rev. D}\ }\textbf {\bibinfo {volume} {108}},\ \bibinfo {pages} {044072} (\bibinfo {year} {2023})},\ \Eprint {https://arxiv.org/abs/2306.02464} {arXiv:2306.02464 [gr-qc]} \BibitemShut {NoStop}%
\bibitem [{\citenamefont {Bonilla}\ \emph {et~al.}(2025)\citenamefont {Bonilla}, \citenamefont {Santoni}, \citenamefont {Nunes},\ and\ \citenamefont {Levi~Said}}]{Bonilla:2025mrt}%
  \BibitemOpen
  \bibfield  {author} {\bibinfo {author} {\bibfnamefont {A.}~\bibnamefont {Bonilla}}, \bibinfo {author} {\bibfnamefont {A.}~\bibnamefont {Santoni}}, \bibinfo {author} {\bibfnamefont {R.~C.}\ \bibnamefont {Nunes}},\ and\ \bibinfo {author} {\bibfnamefont {J.}~\bibnamefont {Levi~Said}},\ }\bibfield  {title} {\bibinfo {title} {{VSL-Gravity in light of PSR B1913+16 full data set: Upper limits on graviton mass and its theoretical consequences}},\ }\href {https://doi.org/10.1016/j.physletb.2025.139388} {\bibfield  {journal} {\bibinfo  {journal} {Phys. Lett. B}\ }\textbf {\bibinfo {volume} {864}},\ \bibinfo {pages} {139388} (\bibinfo {year} {2025})},\ \Eprint {https://arxiv.org/abs/2503.12195} {arXiv:2503.12195 [gr-qc]} \BibitemShut {NoStop}%
\bibitem [{\citenamefont {Bufalo}\ and\ \citenamefont {Cardoso~e Bufalo}(2019)}]{Bufalo:2019kea}%
  \BibitemOpen
  \bibfield  {author} {\bibinfo {author} {\bibfnamefont {R.}~\bibnamefont {Bufalo}}\ and\ \bibinfo {author} {\bibfnamefont {T.}~\bibnamefont {Cardoso~e Bufalo}},\ }\bibfield  {title} {\bibinfo {title} {{Tree-level processes in very special relativity}},\ }\href {https://doi.org/10.1103/PhysRevD.100.125017} {\bibfield  {journal} {\bibinfo  {journal} {Phys. Rev. D}\ }\textbf {\bibinfo {volume} {100}},\ \bibinfo {pages} {125017} (\bibinfo {year} {2019})},\ \Eprint {https://arxiv.org/abs/1911.08386} {arXiv:1911.08386 [hep-th]} \BibitemShut {NoStop}%
\bibitem [{\citenamefont {Alfaro}(2021)}]{PhysRevD.103.075011}%
  \BibitemOpen
  \bibfield  {author} {\bibinfo {author} {\bibfnamefont {J.}~\bibnamefont {Alfaro}},\ }\bibfield  {title} {\bibinfo {title} {Axial anomaly in very special relativity},\ }\href {https://doi.org/10.1103/PhysRevD.103.075011} {\bibfield  {journal} {\bibinfo  {journal} {Phys. Rev. D}\ }\textbf {\bibinfo {volume} {103}},\ \bibinfo {pages} {075011} (\bibinfo {year} {2021})}\BibitemShut {NoStop}%
\bibitem [{\citenamefont {Bufalo}\ \emph {et~al.}(2020)\citenamefont {Bufalo}, \citenamefont {Ghasemkhani},\ and\ \citenamefont {Soto}}]{Bufalo:2020znk}%
  \BibitemOpen
  \bibfield  {author} {\bibinfo {author} {\bibfnamefont {R.}~\bibnamefont {Bufalo}}, \bibinfo {author} {\bibfnamefont {M.}~\bibnamefont {Ghasemkhani}},\ and\ \bibinfo {author} {\bibfnamefont {A.}~\bibnamefont {Soto}},\ }\bibfield  {title} {\bibinfo {title} {{Adler-Bell-Jackiw anomaly in VSR electrodynamics}},\ }\href@noop {} {\  (\bibinfo {year} {2020})},\ \Eprint {https://arxiv.org/abs/2011.10649} {arXiv:2011.10649 [hep-th]} \BibitemShut {NoStop}%
\bibitem [{\citenamefont {Koch}\ \emph {et~al.}(2022)\citenamefont {Koch}, \citenamefont {Mu\~noz},\ and\ \citenamefont {Santoni}}]{Koch:2022jcd}%
  \BibitemOpen
  \bibfield  {author} {\bibinfo {author} {\bibfnamefont {B.}~\bibnamefont {Koch}}, \bibinfo {author} {\bibfnamefont {E.}~\bibnamefont {Mu\~noz}},\ and\ \bibinfo {author} {\bibfnamefont {A.}~\bibnamefont {Santoni}},\ }\bibfield  {title} {\bibinfo {title} {{Corrections to the gyromagnetic factor in very special relativity}},\ }\href {https://doi.org/10.1103/PhysRevD.106.096009} {\bibfield  {journal} {\bibinfo  {journal} {Phys. Rev. D}\ }\textbf {\bibinfo {volume} {106}},\ \bibinfo {pages} {096009} (\bibinfo {year} {2022})},\ \Eprint {https://arxiv.org/abs/2208.09824} {arXiv:2208.09824 [hep-ph]} \BibitemShut {NoStop}%
\bibitem [{\citenamefont {Bufalo}\ and\ \citenamefont {Ghasemkhani}(2019)}]{PhysRevD.100.065024}%
  \BibitemOpen
  \bibfield  {author} {\bibinfo {author} {\bibfnamefont {R.}~\bibnamefont {Bufalo}}\ and\ \bibinfo {author} {\bibfnamefont {M.}~\bibnamefont {Ghasemkhani}},\ }\bibfield  {title} {\bibinfo {title} {Thermal effects of very special relativity quantum electrodynamics},\ }\href {https://doi.org/10.1103/PhysRevD.100.065024} {\bibfield  {journal} {\bibinfo  {journal} {Phys. Rev. D}\ }\textbf {\bibinfo {volume} {100}},\ \bibinfo {pages} {065024} (\bibinfo {year} {2019})}\BibitemShut {NoStop}%
\bibitem [{\citenamefont {Colladay}\ and\ \citenamefont {Kosteleck\'y}(1998)}]{Colladay:1998fq}%
  \BibitemOpen
  \bibfield  {author} {\bibinfo {author} {\bibfnamefont {D.}~\bibnamefont {Colladay}}\ and\ \bibinfo {author} {\bibfnamefont {V.~A.}\ \bibnamefont {Kosteleck\'y}},\ }\bibfield  {title} {\bibinfo {title} {{Lorentz violating extension of the standard model}},\ }\href {https://doi.org/10.1103/PhysRevD.58.116002} {\bibfield  {journal} {\bibinfo  {journal} {Phys. Rev. D}\ }\textbf {\bibinfo {volume} {58}},\ \bibinfo {pages} {116002} (\bibinfo {year} {1998})},\ \Eprint {https://arxiv.org/abs/hep-ph/9809521} {arXiv:hep-ph/9809521} \BibitemShut {NoStop}%
\bibitem [{\citenamefont {Kosteleck\'y}\ and\ \citenamefont {Lehnert}(2001)}]{PhysRevD.63.065008}%
  \BibitemOpen
  \bibfield  {author} {\bibinfo {author} {\bibfnamefont {V.~A.}\ \bibnamefont {Kosteleck\'y}}\ and\ \bibinfo {author} {\bibfnamefont {R.}~\bibnamefont {Lehnert}},\ }\bibfield  {title} {\bibinfo {title} {Stability, causality, and lorentz and $\mathrm{CPT}$ violation},\ }\href {https://doi.org/10.1103/PhysRevD.63.065008} {\bibfield  {journal} {\bibinfo  {journal} {Phys. Rev. D}\ }\textbf {\bibinfo {volume} {63}},\ \bibinfo {pages} {065008} (\bibinfo {year} {2001})}\BibitemShut {NoStop}%
\bibitem [{\citenamefont {Lehnert}(2004)}]{lehnert2004dirac}%
  \BibitemOpen
  \bibfield  {author} {\bibinfo {author} {\bibfnamefont {R.}~\bibnamefont {Lehnert}},\ }\bibfield  {title} {\bibinfo {title} {{Dirac theory within the standard model extension}},\ }\href {https://doi.org/10.1063/1.1769105} {\bibfield  {journal} {\bibinfo  {journal} {J. Math. Phys.}\ }\textbf {\bibinfo {volume} {45}},\ \bibinfo {pages} {3399} (\bibinfo {year} {2004})},\ \Eprint {https://arxiv.org/abs/hep-ph/0401084} {arXiv:hep-ph/0401084} \BibitemShut {NoStop}%
\bibitem [{\citenamefont {Kosteleck\'y}\ and\ \citenamefont {Li}(2021)}]{Kostelecky:2021tdf}%
  \BibitemOpen
  \bibfield  {author} {\bibinfo {author} {\bibfnamefont {V.~A.}\ \bibnamefont {Kosteleck\'y}}\ and\ \bibinfo {author} {\bibfnamefont {Z.}~\bibnamefont {Li}},\ }\bibfield  {title} {\bibinfo {title} {{Searches for beyond-Riemann gravity}},\ }\href {https://doi.org/10.1103/PhysRevD.104.044054} {\bibfield  {journal} {\bibinfo  {journal} {Phys. Rev. D}\ }\textbf {\bibinfo {volume} {104}},\ \bibinfo {pages} {044054} (\bibinfo {year} {2021})},\ \Eprint {https://arxiv.org/abs/2106.11293} {arXiv:2106.11293 [gr-qc]} \BibitemShut {NoStop}%
\bibitem [{\citenamefont {Kosteleck\'y}\ and\ \citenamefont {Russell}(2011)}]{kostelecky2011data}%
  \BibitemOpen
  \bibfield  {author} {\bibinfo {author} {\bibfnamefont {V.~A.}\ \bibnamefont {Kosteleck\'y}}\ and\ \bibinfo {author} {\bibfnamefont {N.}~\bibnamefont {Russell}},\ }\bibfield  {title} {\bibinfo {title} {{Data Tables for Lorentz and CPT Violation}},\ }\href {https://doi.org/10.1103/RevModPhys.83.11} {\bibfield  {journal} {\bibinfo  {journal} {Rev. Mod. Phys.}\ }\textbf {\bibinfo {volume} {83}},\ \bibinfo {pages} {11} (\bibinfo {year} {2011})},\ \Eprint {https://arxiv.org/abs/0801.0287} {arXiv:0801.0287 [hep-ph]} \BibitemShut {NoStop}%
\bibitem [{\citenamefont {Ivanov}\ \emph {et~al.}(2019)\citenamefont {Ivanov}, \citenamefont {Wellenzohn},\ and\ \citenamefont {Abele}}]{Ivanov:2019ouz}%
  \BibitemOpen
  \bibfield  {author} {\bibinfo {author} {\bibfnamefont {A.~N.}\ \bibnamefont {Ivanov}}, \bibinfo {author} {\bibfnamefont {M.}~\bibnamefont {Wellenzohn}},\ and\ \bibinfo {author} {\bibfnamefont {H.}~\bibnamefont {Abele}},\ }\bibfield  {title} {\bibinfo {title} {{Probing of violation of Lorentz invariance by ultracold neutrons in the Standard Model Extension}},\ }\href {https://doi.org/10.1016/j.physletb.2019.134819} {\bibfield  {journal} {\bibinfo  {journal} {Phys. Lett. B}\ }\textbf {\bibinfo {volume} {797}},\ \bibinfo {pages} {134819} (\bibinfo {year} {2019})},\ \Eprint {https://arxiv.org/abs/1908.01498} {arXiv:1908.01498 [hep-ph]} \BibitemShut {NoStop}%
\bibitem [{\citenamefont {Ivanov}\ \emph {et~al.}(2021)\citenamefont {Ivanov}, \citenamefont {Wellenzohn},\ and\ \citenamefont {Abele}}]{Ivanov:2021bvk}%
  \BibitemOpen
  \bibfield  {author} {\bibinfo {author} {\bibfnamefont {A.~N.}\ \bibnamefont {Ivanov}}, \bibinfo {author} {\bibfnamefont {M.}~\bibnamefont {Wellenzohn}},\ and\ \bibinfo {author} {\bibfnamefont {H.}~\bibnamefont {Abele}},\ }\bibfield  {title} {\bibinfo {title} {{Quantum gravitational states of ultracold neutrons as a tool for probing of beyond-Riemann gravity}},\ }\href {https://doi.org/10.1016/j.physletb.2021.136640} {\bibfield  {journal} {\bibinfo  {journal} {Phys. Lett. B}\ }\textbf {\bibinfo {volume} {822}},\ \bibinfo {pages} {136640} (\bibinfo {year} {2021})},\ \Eprint {https://arxiv.org/abs/2109.09982} {arXiv:2109.09982 [gr-qc]} \BibitemShut {NoStop}%
\bibitem [{\citenamefont {Ignatovich}\ and\ \citenamefont {Pontecorvo}(1990)}]{ignatovich1986physics}%
  \BibitemOpen
  \bibfield  {author} {\bibinfo {author} {\bibfnamefont {V.~K.}\ \bibnamefont {Ignatovich}}\ and\ \bibinfo {author} {\bibfnamefont {G.~B.}\ \bibnamefont {Pontecorvo}},\ }\href {https://doi.org/10.1093/oso/9780198510154.001.0001} {\emph {\bibinfo {title} {{The Physics of Ultracold Neutrons}}}}\ (\bibinfo  {publisher} {Oxford University Press},\ \bibinfo {year} {1990})\BibitemShut {NoStop}%
\bibitem [{\citenamefont {Abele}(2008)}]{Abele:2008zz}%
  \BibitemOpen
  \bibfield  {author} {\bibinfo {author} {\bibfnamefont {H.}~\bibnamefont {Abele}},\ }\bibfield  {title} {\bibinfo {title} {{The neutron. Its properties and basic interactions}},\ }\href {https://doi.org/10.1016/j.ppnp.2007.05.002} {\bibfield  {journal} {\bibinfo  {journal} {Prog. Part. Nucl. Phys.}\ }\textbf {\bibinfo {volume} {60}},\ \bibinfo {pages} {1} (\bibinfo {year} {2008})}\BibitemShut {NoStop}%
\bibitem [{\citenamefont {Abele}\ \emph {et~al.}(2011)\citenamefont {Abele}, \citenamefont {Cronenberg}, \citenamefont {Geltenbort}, \citenamefont {Jenke}, \citenamefont {Lins},\ and\ \citenamefont {Saul}}]{abele2011qbounce}%
  \BibitemOpen
  \bibfield  {author} {\bibinfo {author} {\bibfnamefont {H.}~\bibnamefont {Abele}}, \bibinfo {author} {\bibfnamefont {G.}~\bibnamefont {Cronenberg}}, \bibinfo {author} {\bibfnamefont {P.}~\bibnamefont {Geltenbort}}, \bibinfo {author} {\bibfnamefont {T.}~\bibnamefont {Jenke}}, \bibinfo {author} {\bibfnamefont {T.}~\bibnamefont {Lins}},\ and\ \bibinfo {author} {\bibfnamefont {H.}~\bibnamefont {Saul}},\ }\bibfield  {title} {\bibinfo {title} {qbounce, the quantum bouncing ball experiment},\ }\href {https://doi.org/https://doi.org/10.1016/j.phpro.2011.06.011} {\bibfield  {journal} {\bibinfo  {journal} {Physics Procedia}\ }\textbf {\bibinfo {volume} {17}},\ \bibinfo {pages} {4} (\bibinfo {year} {2011})},\ \bibinfo {note} {2nd International Workshop on the Physics of fundamental Symmetries and Interactions - PSI2010}\BibitemShut {NoStop}%
\bibitem [{\citenamefont {Jenke}\ \emph {et~al.}(2009)\citenamefont {Jenke}, \citenamefont {Stadler}, \citenamefont {Abele},\ and\ \citenamefont {Geltenbort}}]{jenke2009q}%
  \BibitemOpen
  \bibfield  {author} {\bibinfo {author} {\bibfnamefont {T.}~\bibnamefont {Jenke}}, \bibinfo {author} {\bibfnamefont {D.}~\bibnamefont {Stadler}}, \bibinfo {author} {\bibfnamefont {H.}~\bibnamefont {Abele}},\ and\ \bibinfo {author} {\bibfnamefont {P.}~\bibnamefont {Geltenbort}},\ }\bibfield  {title} {\bibinfo {title} {Q-bounce—experiments with quantum bouncing ultracold neutrons},\ }\href {https://doi.org/https://doi.org/10.1016/j.nima.2009.07.073} {\bibfield  {journal} {\bibinfo  {journal} {Nuclear Instruments and Methods in Physics Research Section A: Accelerators, Spectrometers, Detectors and Associated Equipment}\ }\textbf {\bibinfo {volume} {611}},\ \bibinfo {pages} {318} (\bibinfo {year} {2009})},\ \bibinfo {note} {particle Physics with Slow Neutrons}\BibitemShut {NoStop}%
\bibitem [{\citenamefont {Abele}\ and\ \citenamefont {Leeb}(2012)}]{Abele:2012dn}%
  \BibitemOpen
  \bibfield  {author} {\bibinfo {author} {\bibfnamefont {H.}~\bibnamefont {Abele}}\ and\ \bibinfo {author} {\bibfnamefont {H.}~\bibnamefont {Leeb}},\ }\bibfield  {title} {\bibinfo {title} {{Gravitation and quantum interference experiments with neutrons}},\ }\href {https://doi.org/10.1088/1367-2630/14/5/055010} {\bibfield  {journal} {\bibinfo  {journal} {New J. Phys.}\ }\textbf {\bibinfo {volume} {14}},\ \bibinfo {pages} {055010} (\bibinfo {year} {2012})},\ \Eprint {https://arxiv.org/abs/1207.2953} {arXiv:1207.2953 [hep-ph]} \BibitemShut {NoStop}%
\bibitem [{\citenamefont {Pitschmann}\ and\ \citenamefont {Abele}(2019)}]{pitschmann2019schr}%
  \BibitemOpen
  \bibfield  {author} {\bibinfo {author} {\bibfnamefont {M.}~\bibnamefont {Pitschmann}}\ and\ \bibinfo {author} {\bibfnamefont {H.}~\bibnamefont {Abele}},\ }\bibfield  {title} {\bibinfo {title} {{Schr\"odinger Equation for a Non-Relativistic Particle in a Gravitational Field confined by Two Vibrating Mirrors}},\ }\href@noop {} {\  (\bibinfo {year} {2019})},\ \Eprint {https://arxiv.org/abs/1912.12236} {arXiv:1912.12236 [quant-ph]} \BibitemShut {NoStop}%
\bibitem [{\citenamefont {Santoni}(2024)}]{Santoni:2024coa}%
  \BibitemOpen
  \bibfield  {author} {\bibinfo {author} {\bibfnamefont {A.}~\bibnamefont {Santoni}},\ }\emph {\bibinfo {title} {{Delving into the phenomenology of very special relativity: From subatomic particles to binary stars}}},\ \href {https://doi.org/10.34726/hss.2024.124925} {Ph.D. thesis},\ \bibinfo  {school} {Vienna, Tech. U.} (\bibinfo {year} {2024}),\ \Eprint {https://arxiv.org/abs/2409.03104} {arXiv:2409.03104 [hep-ph]} \BibitemShut {NoStop}%
\bibitem [{\citenamefont {Bender}(2007)}]{Bender:2007nj}%
  \BibitemOpen
  \bibfield  {author} {\bibinfo {author} {\bibfnamefont {C.~M.}\ \bibnamefont {Bender}},\ }\bibfield  {title} {\bibinfo {title} {{Making sense of non-Hermitian Hamiltonians}},\ }\href {https://doi.org/10.1088/0034-4885/70/6/R03} {\bibfield  {journal} {\bibinfo  {journal} {Rept. Prog. Phys.}\ }\textbf {\bibinfo {volume} {70}},\ \bibinfo {pages} {947} (\bibinfo {year} {2007})},\ \Eprint {https://arxiv.org/abs/hep-th/0703096} {arXiv:hep-th/0703096} \BibitemShut {NoStop}%
\bibitem [{\citenamefont {Alexandre}\ and\ \citenamefont {Bender}(2015)}]{alexandre2015foldy}%
  \BibitemOpen
  \bibfield  {author} {\bibinfo {author} {\bibfnamefont {J.}~\bibnamefont {Alexandre}}\ and\ \bibinfo {author} {\bibfnamefont {C.~M.}\ \bibnamefont {Bender}},\ }\bibfield  {title} {\bibinfo {title} {Foldy--wouthuysen transformation for non-hermitian hamiltonians},\ }\href {https://doi.org/10.1088/1742-6596/631/1/012071} {\bibfield  {journal} {\bibinfo  {journal} {Journal of Physics A: Mathematical and Theoretical}\ }\textbf {\bibinfo {volume} {48}},\ \bibinfo {pages} {185403} (\bibinfo {year} {2015})}\BibitemShut {NoStop}%
\bibitem [{\citenamefont {Bender}\ and\ \citenamefont {Hook}(2023)}]{Bender:2023cem}%
  \BibitemOpen
  \bibfield  {author} {\bibinfo {author} {\bibfnamefont {C.~M.}\ \bibnamefont {Bender}}\ and\ \bibinfo {author} {\bibfnamefont {D.~W.}\ \bibnamefont {Hook}},\ }\bibfield  {title} {\bibinfo {title} {{PT-symmetric quantum mechanics}},\ }\href@noop {} {\  (\bibinfo {year} {2023})},\ \Eprint {https://arxiv.org/abs/2312.17386} {arXiv:2312.17386 [quant-ph]} \BibitemShut {NoStop}%
\bibitem [{\citenamefont {Bender}(2020)}]{Bender:2020gbh}%
  \BibitemOpen
  \bibfield  {author} {\bibinfo {author} {\bibfnamefont {C.~M.}\ \bibnamefont {Bender}},\ }\bibfield  {title} {\bibinfo {title} {{$\mathcal{PT}$-symmetric quantum field theory}},\ }\href {https://doi.org/10.1088/1742-6596/1586/1/012004} {\bibfield  {journal} {\bibinfo  {journal} {J. Phys. Conf. Ser.}\ }\textbf {\bibinfo {volume} {1586}},\ \bibinfo {pages} {012004} (\bibinfo {year} {2020})}\BibitemShut {NoStop}%
\bibitem [{\citenamefont {Bender}\ \emph {et~al.}(2005)\citenamefont {Bender}, \citenamefont {Brandt}, \citenamefont {Chen},\ and\ \citenamefont {Wang}}]{PhysRevD.71.025014}%
  \BibitemOpen
  \bibfield  {author} {\bibinfo {author} {\bibfnamefont {C.~M.}\ \bibnamefont {Bender}}, \bibinfo {author} {\bibfnamefont {S.~F.}\ \bibnamefont {Brandt}}, \bibinfo {author} {\bibfnamefont {J.-H.}\ \bibnamefont {Chen}},\ and\ \bibinfo {author} {\bibfnamefont {Q.}~\bibnamefont {Wang}},\ }\bibfield  {title} {\bibinfo {title} {Ghost busting: $\mathcal{P}\mathcal{T}$-symmetric interpretation of the lee model},\ }\href {https://doi.org/10.1103/PhysRevD.71.025014} {\bibfield  {journal} {\bibinfo  {journal} {Phys. Rev. D}\ }\textbf {\bibinfo {volume} {71}},\ \bibinfo {pages} {025014} (\bibinfo {year} {2005})}\BibitemShut {NoStop}%
\bibitem [{\citenamefont {Giacosa}(2020)}]{Giacosa:2020tha}%
  \BibitemOpen
  \bibfield  {author} {\bibinfo {author} {\bibfnamefont {F.}~\bibnamefont {Giacosa}},\ }\bibfield  {title} {\bibinfo {title} {{The Lee model: a tool to study decays}},\ }\href {https://doi.org/10.1088/1742-6596/1612/1/012012} {\bibfield  {journal} {\bibinfo  {journal} {J. Phys. Conf. Ser.}\ }\textbf {\bibinfo {volume} {1612}},\ \bibinfo {pages} {012012} (\bibinfo {year} {2020})},\ \Eprint {https://arxiv.org/abs/2001.07781} {arXiv:2001.07781 [hep-ph]} \BibitemShut {NoStop}%
\bibitem [{\citenamefont {Ilderton}(2016)}]{Ilderton:2016rqk}%
  \BibitemOpen
  \bibfield  {author} {\bibinfo {author} {\bibfnamefont {A.}~\bibnamefont {Ilderton}},\ }\bibfield  {title} {\bibinfo {title} {{Very Special Relativity as a background field theory}},\ }\href {https://doi.org/10.1103/PhysRevD.94.045019} {\bibfield  {journal} {\bibinfo  {journal} {Phys. Rev. D}\ }\textbf {\bibinfo {volume} {94}},\ \bibinfo {pages} {045019} (\bibinfo {year} {2016})},\ \Eprint {https://arxiv.org/abs/1605.04967} {arXiv:1605.04967 [hep-th]} \BibitemShut {NoStop}%
\bibitem [{\citenamefont {De~Oliveira}\ and\ \citenamefont {Tiomno}(1962)}]{de1962representations}%
  \BibitemOpen
  \bibfield  {author} {\bibinfo {author} {\bibfnamefont {C.}~\bibnamefont {De~Oliveira}}\ and\ \bibinfo {author} {\bibfnamefont {J.}~\bibnamefont {Tiomno}},\ }\bibfield  {title} {\bibinfo {title} {Representations of dirac equation in general relativity},\ }\href {https://doi.org/https://doi.org/10.1007/BF02816716} {\bibfield  {journal} {\bibinfo  {journal} {Il Nuovo Cimento (1955-1965)}\ }\textbf {\bibinfo {volume} {24}},\ \bibinfo {pages} {672} (\bibinfo {year} {1962})}\BibitemShut {NoStop}%
\bibitem [{\citenamefont {Pollock}(2010)}]{Pollock:2010zz}%
  \BibitemOpen
  \bibfield  {author} {\bibinfo {author} {\bibfnamefont {M.~D.}\ \bibnamefont {Pollock}},\ }\bibfield  {title} {\bibinfo {title} {{On the Dirac equation in curved space-time}},\ }\href {www.actaphys.uj.edu.pl/fulltext?series=Reg&vol=41&page=1827} {\bibfield  {journal} {\bibinfo  {journal} {Acta Phys. Polon. B}\ }\textbf {\bibinfo {volume} {41}},\ \bibinfo {pages} {1827} (\bibinfo {year} {2010})}\BibitemShut {NoStop}%
\bibitem [{\citenamefont {Weinberg}(1972)}]{weinberg1972gravitation}%
  \BibitemOpen
  \bibfield  {author} {\bibinfo {author} {\bibfnamefont {S.}~\bibnamefont {Weinberg}},\ }\href {www.archive.org/details/WeinbergS.GravitationAndCosmology..PrinciplesAndApplicationsOfTheGeneralTheoryOf} {\emph {\bibinfo {title} {{Gravitation and Cosmology}: {Principles and Applications of the General Theory of Relativity}}}}\ (\bibinfo  {publisher} {John Wiley and Sons},\ \bibinfo {address} {New York},\ \bibinfo {year} {1972})\BibitemShut {NoStop}%
\bibitem [{\citenamefont {Huang}(2005)}]{Huang:2005rn}%
  \BibitemOpen
  \bibfield  {author} {\bibinfo {author} {\bibfnamefont {X.-B.}\ \bibnamefont {Huang}},\ }\bibfield  {title} {\bibinfo {title} {{Exact solutions of the Dirac equation in Robertson-Walker space-time}},\ }\href@noop {} {\  (\bibinfo {year} {2005})},\ \Eprint {https://arxiv.org/abs/gr-qc/0501077} {arXiv:gr-qc/0501077} \BibitemShut {NoStop}%
\bibitem [{\citenamefont {Parker}(1980)}]{parker1980one}%
  \BibitemOpen
  \bibfield  {author} {\bibinfo {author} {\bibfnamefont {L.}~\bibnamefont {Parker}},\ }\bibfield  {title} {\bibinfo {title} {{One-electron atom as a probe of space-time curvature}},\ }\href {https://doi.org/10.1103/PhysRevD.22.1922} {\bibfield  {journal} {\bibinfo  {journal} {Phys. Rev. D}\ }\textbf {\bibinfo {volume} {22}},\ \bibinfo {pages} {1922} (\bibinfo {year} {1980})}\BibitemShut {NoStop}%
\bibitem [{\citenamefont {Huang}\ and\ \citenamefont {Parker}(2009)}]{huang2009hermiticity}%
  \BibitemOpen
  \bibfield  {author} {\bibinfo {author} {\bibfnamefont {X.}~\bibnamefont {Huang}}\ and\ \bibinfo {author} {\bibfnamefont {L.}~\bibnamefont {Parker}},\ }\bibfield  {title} {\bibinfo {title} {{Hermiticity of the Dirac Hamiltonian in Curved Spacetime}},\ }\href {https://doi.org/10.1103/PhysRevD.79.024020} {\bibfield  {journal} {\bibinfo  {journal} {Phys. Rev. D}\ }\textbf {\bibinfo {volume} {79}},\ \bibinfo {pages} {024020} (\bibinfo {year} {2009})},\ \Eprint {https://arxiv.org/abs/0811.2296} {arXiv:0811.2296 [hep-th]} \BibitemShut {NoStop}%
\bibitem [{\citenamefont {Misner}\ \emph {et~al.}(1973)\citenamefont {Misner}, \citenamefont {Thorne},\ and\ \citenamefont {Wheeler}}]{Misner:1973prb}%
  \BibitemOpen
  \bibfield  {author} {\bibinfo {author} {\bibfnamefont {C.~W.}\ \bibnamefont {Misner}}, \bibinfo {author} {\bibfnamefont {K.~S.}\ \bibnamefont {Thorne}},\ and\ \bibinfo {author} {\bibfnamefont {J.~A.}\ \bibnamefont {Wheeler}},\ }\href {www.press.princeton.edu/books/hardcover/9780691177793/gravitation} {\emph {\bibinfo {title} {{Gravitation}}}}\ (\bibinfo  {publisher} {W. H. Freeman},\ \bibinfo {address} {San Francisco},\ \bibinfo {year} {1973})\BibitemShut {NoStop}%
\bibitem [{\citenamefont {Rindler}(1969)}]{Rindler1969-RINER}%
  \BibitemOpen
  \bibfield  {author} {\bibinfo {author} {\bibfnamefont {W.}~\bibnamefont {Rindler}},\ }\href {https://doi.org/https://doi.org/10.1007/978-3-642-86650-0} {\emph {\bibinfo {title} {Essential Relativity}}}\ (\bibinfo  {publisher} {Van Nostrand Reinhold Co.},\ \bibinfo {address} {New York,},\ \bibinfo {year} {1969})\BibitemShut {NoStop}%
\bibitem [{\citenamefont {M{\o}ller}(1943)}]{moller1943homogeneous}%
  \BibitemOpen
  \bibfield  {author} {\bibinfo {author} {\bibfnamefont {C.}~\bibnamefont {M{\o}ller}},\ }\href {http://publ.royalacademy.dk/backend/web/uploads/2019-07-23/AFL\%202/M_20_00_00_1942-1943_6345/M_20_19_00_1943_1035.pdf} {\emph {\bibinfo {title} {On homogeneous gravitational fields in the general theory of relativity and the clock paradox}}}\ (\bibinfo  {publisher} {Munksgaard},\ \bibinfo {year} {1943})\BibitemShut {NoStop}%
\bibitem [{\citenamefont {Maluf}\ and\ \citenamefont {Faria}(2008)}]{maluf2008construction}%
  \BibitemOpen
  \bibfield  {author} {\bibinfo {author} {\bibfnamefont {J.~W.}\ \bibnamefont {Maluf}}\ and\ \bibinfo {author} {\bibfnamefont {F.~F.}\ \bibnamefont {Faria}},\ }\bibfield  {title} {\bibinfo {title} {{On the construction of Fermi-Walker transported frames}},\ }\href {https://doi.org/10.1002/andp.200810289} {\bibfield  {journal} {\bibinfo  {journal} {Annalen Phys.}\ }\textbf {\bibinfo {volume} {17}},\ \bibinfo {pages} {326} (\bibinfo {year} {2008})},\ \Eprint {https://arxiv.org/abs/0804.2502} {arXiv:0804.2502 [gr-qc]} \BibitemShut {NoStop}%
\bibitem [{\citenamefont {Klein}\ and\ \citenamefont {Collas}(2008)}]{klein2008general}%
  \BibitemOpen
  \bibfield  {author} {\bibinfo {author} {\bibfnamefont {D.}~\bibnamefont {Klein}}\ and\ \bibinfo {author} {\bibfnamefont {P.}~\bibnamefont {Collas}},\ }\bibfield  {title} {\bibinfo {title} {{General Transformation Formulas for Fermi-Walker Coordinates}},\ }\href {https://doi.org/10.1088/0264-9381/25/14/145019} {\bibfield  {journal} {\bibinfo  {journal} {Class. Quant. Grav.}\ }\textbf {\bibinfo {volume} {25}},\ \bibinfo {pages} {145019} (\bibinfo {year} {2008})},\ \Eprint {https://arxiv.org/abs/0712.3838} {arXiv:0712.3838 [gr-qc]} \BibitemShut {NoStop}%
\bibitem [{\citenamefont {Bluhm}(2015)}]{PhysRevD.91.065034}%
  \BibitemOpen
  \bibfield  {author} {\bibinfo {author} {\bibfnamefont {R.}~\bibnamefont {Bluhm}},\ }\bibfield  {title} {\bibinfo {title} {Explicit versus spontaneous diffeomorphism breaking in gravity},\ }\href {https://doi.org/10.1103/PhysRevD.91.065034} {\bibfield  {journal} {\bibinfo  {journal} {Phys. Rev. D}\ }\textbf {\bibinfo {volume} {91}},\ \bibinfo {pages} {065034} (\bibinfo {year} {2015})}\BibitemShut {NoStop}%
\bibitem [{\citenamefont {Ju}\ \emph {et~al.}(2022)\citenamefont {Ju}, \citenamefont {Miranowicz}, \citenamefont {Minganti}, \citenamefont {Chan}, \citenamefont {Chen},\ and\ \citenamefont {Nori}}]{Ju:2021vvs}%
  \BibitemOpen
  \bibfield  {author} {\bibinfo {author} {\bibfnamefont {C.-Y.}\ \bibnamefont {Ju}}, \bibinfo {author} {\bibfnamefont {A.}~\bibnamefont {Miranowicz}}, \bibinfo {author} {\bibfnamefont {F.}~\bibnamefont {Minganti}}, \bibinfo {author} {\bibfnamefont {C.-T.}\ \bibnamefont {Chan}}, \bibinfo {author} {\bibfnamefont {G.-Y.}\ \bibnamefont {Chen}},\ and\ \bibinfo {author} {\bibfnamefont {F.}~\bibnamefont {Nori}},\ }\bibfield  {title} {\bibinfo {title} {{Einstein's quantum elevator: Hermitization of non-Hermitian Hamiltonians via a generalized vielbein formalism}},\ }\href {https://doi.org/10.1103/PhysRevResearch.4.023070} {\bibfield  {journal} {\bibinfo  {journal} {Phys. Rev. Res.}\ }\textbf {\bibinfo {volume} {4}},\ \bibinfo {pages} {023070} (\bibinfo {year} {2022})},\ \Eprint {https://arxiv.org/abs/2107.11910} {arXiv:2107.11910 [quant-ph]} \BibitemShut {NoStop}%
\bibitem [{\citenamefont {Obukhov}\ \emph {et~al.}(2011)\citenamefont {Obukhov}, \citenamefont {Silenko},\ and\ \citenamefont {Teryaev}}]{Obukhov:2011ks}%
  \BibitemOpen
  \bibfield  {author} {\bibinfo {author} {\bibfnamefont {Y.~N.}\ \bibnamefont {Obukhov}}, \bibinfo {author} {\bibfnamefont {A.~J.}\ \bibnamefont {Silenko}},\ and\ \bibinfo {author} {\bibfnamefont {O.~V.}\ \bibnamefont {Teryaev}},\ }\bibfield  {title} {\bibinfo {title} {{Dirac fermions in strong gravitational fields}},\ }\href {https://doi.org/10.1103/PhysRevD.84.024025} {\bibfield  {journal} {\bibinfo  {journal} {Phys. Rev. D}\ }\textbf {\bibinfo {volume} {84}},\ \bibinfo {pages} {024025} (\bibinfo {year} {2011})},\ \Eprint {https://arxiv.org/abs/1106.0173} {arXiv:1106.0173 [hep-th]} \BibitemShut {NoStop}%
\bibitem [{\citenamefont {Arminjon}(2006)}]{arminjon2006post}%
  \BibitemOpen
  \bibfield  {author} {\bibinfo {author} {\bibfnamefont {M.}~\bibnamefont {Arminjon}},\ }\bibfield  {title} {\bibinfo {title} {{Post-Newtonian equation for the energy levels of a Dirac particle in a static metric}},\ }\href {https://doi.org/10.1103/PhysRevD.74.065017} {\bibfield  {journal} {\bibinfo  {journal} {Phys. Rev. D}\ }\textbf {\bibinfo {volume} {74}},\ \bibinfo {pages} {065017} (\bibinfo {year} {2006})},\ \Eprint {https://arxiv.org/abs/gr-qc/0606036} {arXiv:gr-qc/0606036} \BibitemShut {NoStop}%
\bibitem [{\citenamefont {Foldy}\ and\ \citenamefont {Wouthuysen}(1950)}]{foldy1950dirac}%
  \BibitemOpen
  \bibfield  {author} {\bibinfo {author} {\bibfnamefont {L.~L.}\ \bibnamefont {Foldy}}\ and\ \bibinfo {author} {\bibfnamefont {S.~A.}\ \bibnamefont {Wouthuysen}},\ }\bibfield  {title} {\bibinfo {title} {On the dirac theory of spin 1/2 particles and its non-relativistic limit},\ }\href {https://doi.org/10.1103/PhysRev.78.29} {\bibfield  {journal} {\bibinfo  {journal} {Phys. Rev.}\ }\textbf {\bibinfo {volume} {78}},\ \bibinfo {pages} {29} (\bibinfo {year} {1950})}\BibitemShut {NoStop}%
\bibitem [{\citenamefont {Bjorken}\ and\ \citenamefont {Drell}(1965)}]{Bjorken:1965sts}%
  \BibitemOpen
  \bibfield  {author} {\bibinfo {author} {\bibfnamefont {J.~D.}\ \bibnamefont {Bjorken}}\ and\ \bibinfo {author} {\bibfnamefont {S.~D.}\ \bibnamefont {Drell}},\ }\href {https://doi.org/https://doi.org/10.1063/1.3047288} {\emph {\bibinfo {title} {{Relativistic Quantum Mechanics}}}},\ International Series In Pure and Applied Physics\ (\bibinfo  {publisher} {McGraw-Hill},\ \bibinfo {address} {New York},\ \bibinfo {year} {1965})\BibitemShut {NoStop}%
\bibitem [{\citenamefont {Berestetskii}\ \emph {et~al.}(1982)\citenamefont {Berestetskii}, \citenamefont {Lifshitz},\ and\ \citenamefont {Pitaevskii}}]{Berestetskii:1982qgu}%
  \BibitemOpen
  \bibfield  {author} {\bibinfo {author} {\bibfnamefont {V.~B.}\ \bibnamefont {Berestetskii}}, \bibinfo {author} {\bibfnamefont {E.~M.}\ \bibnamefont {Lifshitz}},\ and\ \bibinfo {author} {\bibfnamefont {L.~P.}\ \bibnamefont {Pitaevskii}},\ }\href {https://doi.org/https://doi.org/10.1016/C2009-0-24486-2} {\emph {\bibinfo {title} {{Quantum Electrodynamics}}}},\ \bibinfo {series} {Course of Theoretical Physics}, Vol.~\bibinfo {volume} {4}\ (\bibinfo  {publisher} {Pergamon Press},\ \bibinfo {address} {Oxford},\ \bibinfo {year} {1982})\BibitemShut {NoStop}%
\bibitem [{\citenamefont {Chandrasekhar}(1965)}]{chandrasekhar1965post}%
  \BibitemOpen
  \bibfield  {author} {\bibinfo {author} {\bibfnamefont {S.}~\bibnamefont {Chandrasekhar}},\ }\bibfield  {title} {\bibinfo {title} {The post-newtonian equations of hydrodynamics in general relativity},\ }\href {https://doi.org/https://ui.adsabs.harvard.edu/link_gateway/1965ApJ...142.1488C/doi:10.1086/148432} {\bibfield  {journal} {\bibinfo  {journal} {Astrophysical Journal}\ }\textbf {\bibinfo {volume} {142}},\ \bibinfo {pages} {1488} (\bibinfo {year} {1965})}\BibitemShut {NoStop}%
\bibitem [{\citenamefont {Nelson}(1990)}]{nelson1990post}%
  \BibitemOpen
  \bibfield  {author} {\bibinfo {author} {\bibfnamefont {R.~A.}\ \bibnamefont {Nelson}},\ }\bibfield  {title} {\bibinfo {title} {Post-newtonian approximation for an accelerated, rotating frame of reference},\ }\href {https://doi.org/10.1007/BF00756150} {\bibfield  {journal} {\bibinfo  {journal} {General relativity and gravitation}\ }\textbf {\bibinfo {volume} {22}},\ \bibinfo {pages} {431} (\bibinfo {year} {1990})}\BibitemShut {NoStop}%
\bibitem [{\citenamefont {Bern}\ \emph {et~al.}(2019)\citenamefont {Bern}, \citenamefont {Cheung}, \citenamefont {Roiban}, \citenamefont {Shen}, \citenamefont {Solon},\ and\ \citenamefont {Zeng}}]{bern2019black}%
  \BibitemOpen
  \bibfield  {author} {\bibinfo {author} {\bibfnamefont {Z.}~\bibnamefont {Bern}}, \bibinfo {author} {\bibfnamefont {C.}~\bibnamefont {Cheung}}, \bibinfo {author} {\bibfnamefont {R.}~\bibnamefont {Roiban}}, \bibinfo {author} {\bibfnamefont {C.-H.}\ \bibnamefont {Shen}}, \bibinfo {author} {\bibfnamefont {M.~P.}\ \bibnamefont {Solon}},\ and\ \bibinfo {author} {\bibfnamefont {M.}~\bibnamefont {Zeng}},\ }\bibfield  {title} {\bibinfo {title} {{Black Hole Binary Dynamics from the Double Copy and Effective Theory}},\ }\href {https://doi.org/10.1007/JHEP10(2019)206} {\bibfield  {journal} {\bibinfo  {journal} {JHEP}\ }\textbf {\bibinfo {volume} {10}},\ \bibinfo {pages} {206}},\ \Eprint {https://arxiv.org/abs/1908.01493} {arXiv:1908.01493 [hep-th]} \BibitemShut {NoStop}%
\bibitem [{\citenamefont {Koch}\ \emph {et~al.}(2024)\citenamefont {Koch}, \citenamefont {Mu\~noz},\ and\ \citenamefont {Santoni}}]{PhysRevD.109.064085}%
  \BibitemOpen
  \bibfield  {author} {\bibinfo {author} {\bibfnamefont {B.}~\bibnamefont {Koch}}, \bibinfo {author} {\bibfnamefont {E.}~\bibnamefont {Mu\~noz}},\ and\ \bibinfo {author} {\bibfnamefont {A.}~\bibnamefont {Santoni}},\ }\bibfield  {title} {\bibinfo {title} {Ultracold neutrons in the low curvature limit: Remarks on the post-newtonian effects},\ }\href {https://doi.org/10.1103/PhysRevD.109.064085} {\bibfield  {journal} {\bibinfo  {journal} {Phys. Rev. D}\ }\textbf {\bibinfo {volume} {109}},\ \bibinfo {pages} {064085} (\bibinfo {year} {2024})}\BibitemShut {NoStop}%
\bibitem [{\citenamefont {Silenko}(2025)}]{Silenko:2024zkt}%
  \BibitemOpen
  \bibfield  {author} {\bibinfo {author} {\bibfnamefont {A.~J.}\ \bibnamefont {Silenko}},\ }\bibfield  {title} {\bibinfo {title} {{Leading correction to the relativistic Foldy-Wouthuysen Hamiltonian}},\ }\href {https://doi.org/10.1103/PhysRevA.111.032210} {\bibfield  {journal} {\bibinfo  {journal} {Phys. Rev. A}\ }\textbf {\bibinfo {volume} {111}},\ \bibinfo {pages} {032210} (\bibinfo {year} {2025})},\ \Eprint {https://arxiv.org/abs/2408.01770} {arXiv:2408.01770 [quant-ph]} \BibitemShut {NoStop}%
\bibitem [{\citenamefont {Neznamov}\ and\ \citenamefont {Silenko}(2009)}]{neznamov2009foldy}%
  \BibitemOpen
  \bibfield  {author} {\bibinfo {author} {\bibfnamefont {V.}~\bibnamefont {Neznamov}}\ and\ \bibinfo {author} {\bibfnamefont {A.}~\bibnamefont {Silenko}},\ }\bibfield  {title} {\bibinfo {title} {Foldy--wouthuysen wave functions and conditions of transformation between dirac and foldy--wouthuysen representations},\ }\href@noop {} {\bibfield  {journal} {\bibinfo  {journal} {Journal of mathematical physics}\ }\textbf {\bibinfo {volume} {50}} (\bibinfo {year} {2009})}\BibitemShut {NoStop}%
\bibitem [{\citenamefont {Silenko}(2009)}]{Silenko:2009if}%
  \BibitemOpen
  \bibfield  {author} {\bibinfo {author} {\bibfnamefont {A.~J.}\ \bibnamefont {Silenko}},\ }\bibfield  {title} {\bibinfo {title} {{Foldy-Wouthuysen Transformation and Semiclassical Transition for Relativistic Quantum Mechanics}},\ }in\ \href@noop {} {\emph {\bibinfo {booktitle} {{16th International Congress on Mathematical Physics}}}}\ (\bibinfo {year} {2009})\ \Eprint {https://arxiv.org/abs/0910.5155} {arXiv:0910.5155 [hep-th]} \BibitemShut {NoStop}%
\bibitem [{\citenamefont {Mohr}\ \emph {et~al.}(2024)\citenamefont {Mohr}, \citenamefont {Newell}, \citenamefont {Taylor},\ and\ \citenamefont {Tiesinga}}]{Mohr:2024kco}%
  \BibitemOpen
  \bibfield  {author} {\bibinfo {author} {\bibfnamefont {P.}~\bibnamefont {Mohr}}, \bibinfo {author} {\bibfnamefont {D.}~\bibnamefont {Newell}}, \bibinfo {author} {\bibfnamefont {B.}~\bibnamefont {Taylor}},\ and\ \bibinfo {author} {\bibfnamefont {E.}~\bibnamefont {Tiesinga}},\ }\bibfield  {title} {\bibinfo {title} {{CODATA Recommended Values of the Fundamental Physical Constants: 2022}},\ }\href@noop {} {\bibfield  {journal} {\bibinfo  {journal} {arXiv preprint}\ } (\bibinfo {year} {2024})},\ \Eprint {https://arxiv.org/abs/2409.03787} {arXiv:2409.03787 [hep-ph]} \BibitemShut {NoStop}%
\bibitem [{\citenamefont {Micko}\ \emph {et~al.}(2023)\citenamefont {Micko}, \citenamefont {Di~Pumpo}, \citenamefont {Bosina}, \citenamefont {Cranganore}, \citenamefont {Jenke}, \citenamefont {Pitschmann}, \citenamefont {Roccia}, \citenamefont {Sedmik},\ and\ \citenamefont {Abele}}]{Micko:2023oar}%
  \BibitemOpen
  \bibfield  {author} {\bibinfo {author} {\bibfnamefont {J.}~\bibnamefont {Micko}}, \bibinfo {author} {\bibfnamefont {F.}~\bibnamefont {Di~Pumpo}}, \bibinfo {author} {\bibfnamefont {J.}~\bibnamefont {Bosina}}, \bibinfo {author} {\bibfnamefont {S.~S.}\ \bibnamefont {Cranganore}}, \bibinfo {author} {\bibfnamefont {T.}~\bibnamefont {Jenke}}, \bibinfo {author} {\bibfnamefont {M.}~\bibnamefont {Pitschmann}}, \bibinfo {author} {\bibfnamefont {S.}~\bibnamefont {Roccia}}, \bibinfo {author} {\bibfnamefont {R.~I.~P.}\ \bibnamefont {Sedmik}},\ and\ \bibinfo {author} {\bibfnamefont {H.}~\bibnamefont {Abele}},\ }\bibfield  {title} {\bibinfo {title} {{UGR tests with atomic clocks and atom interferometers}},\ }in\ \href@noop {} {\emph {\bibinfo {booktitle} {{56th Rencontres de Moriond on Gravitation}}}}\ (\bibinfo {year} {2023})\ \Eprint {https://arxiv.org/abs/2203.16375} {arXiv:2203.16375 [gr-qc]} \BibitemShut {NoStop}%
\end{thebibliography}%

\end{document}